\documentclass[journal]{IEEEtran}

\usepackage[T1]{fontenc}
\usepackage[utf8]{inputenc}
\usepackage{cite}
\usepackage{amsmath,amssymb}
\usepackage{graphicx}
\usepackage{booktabs}
\usepackage{multirow}
\usepackage{array}
\usepackage{tabularx}
\usepackage{url}
\usepackage{hyperref}
\usepackage[table]{xcolor}
\usepackage{microtype}
\usepackage{siunitx}
\usepackage{pifont}
\usepackage{tikz}

\definecolor{orcidgreen}{HTML}{A6CE39}
\newcommand{\orcidmark}{%
  \begin{tikzpicture}[baseline=-0.45ex,scale=1.15]
    \fill[orcidgreen] (0,0) circle (0.55ex);
    \fill[white] (-0.28ex,-0.34ex) rectangle (-0.13ex,0.13ex);
    \fill[white] (-0.205ex,0.26ex) circle (0.085ex);
    \draw[white,line width=0.15ex] (0.06ex,-0.27ex) -- (0.06ex,0.2ex);
    \draw[white,line width=0.15ex] (0.06ex,0.2ex)
      arc (90:-90:0.235ex and 0.235ex);
  \end{tikzpicture}%
}
\newcommand{\orcid}[1]{\href{https://orcid.org/#1}{\orcidmark}}

\graphicspath{{./}}

\newcommand{\mlkem}{\mbox{ML-KEM}}
\newcommand{\mldsa}{\mbox{ML-DSA}}
\newcommand{\armmo}{\mbox{Cortex-M0+}}
\newcommand{\armfour}{\mbox{Cortex-M4}}
\newcommand{\cmark}{\ding{51}}

\begin{document}

\title{Benchmarking NIST-Standardised ML-KEM and ML-DSA on ARM Cortex-M0+: Latency, Rejection-Sampling Variance, and Memory on the RP2040}

\author{%
\IEEEauthorblockN{Rojin Chhetri\,\orcid{0009-0005-2757-3436}, Sijan Dhakal, Asmita Gautam}\\
\IEEEauthorblockA{\textit{rojin@softora.edu.np, sijan@softora.edu.np, asmita@softora.edu.np}}
}

\maketitle

\begin{abstract}
Internet of Things devices with 10--20\,year lifespans need
post-quantum migration, yet the finalised NIST standards remain
sparsely benchmarked on the most constrained 32-bit ARM class.
We present, to our knowledge, the first isolated algorithm-level
benchmarks of ML-KEM (FIPS~203) and ML-DSA (FIPS~204) on
ARM~\armmo{}: all six parameter sets on the RP2040
(Raspberry~Pi~Pico) at \SI{125}{\mega\hertz}, using PQClean
reference~C.
A full ML-KEM-512 key exchange completes in \SI{35.7}{\milli\second},
17$\times$ faster than our mbedTLS ECDH\,P-256 baseline on the same
hardware---a comparison of two libraries, not two algorithms.
The RP2040's single-cycle multiplier makes these figures optimistic
for \armmo{} parts with a multi-cycle multiplier; we estimate that
penalty at ${\approx}1.5\times$ by instruction count, the
multiply-free Keccak layer dominating reference~C on this core,
which would place the same handshake at
${\approx}\SI{54}{\milli\second}$ and still ${\approx}11\times$
ahead of the ECDH baseline on 32-cycle-multiplier silicon.
ML-DSA signing varies widely under rejection sampling
(coefficient of variation 66.0--73.5\%), with a tail no finite
sample bounds.  At ML-DSA-44, the one level where we ran
\num{10000} signatures, $p_{99}$ is \SI{584}{\milli\second} and the
observed maximum \SI{1157}{\milli\second}---respectively
\SI{19}{\percent} and $2.1\times$ above what the $n{=}100$ campaign
routine in embedded benchmarking reports.  Our ML-DSA-65 and~-87
quantiles come from $n{=}100$ and are therefore lower bounds
($p_{99}\ge\SI{1125}{\milli\second}$ at ML-DSA-87).
We further recover the rejection-sampling iteration count of
individual signatures from wall-clock time alone, identifying the
otherwise degenerate timing-lattice fit by requiring recovered
counts to follow the geometric law rejection sampling induces.  We
analyse this as a timing side channel: it recovers the \emph{count}
of rejected attempts, not their \emph{content}, and we claim no key
recovery.
The \armmo{} incurs only a ${\sim}2.0\times$ slowdown
(\numrange{2.01}{2.04}$\times$ cycle-matched at \texttt{-O3};
\numrange{1.96}{2.09}$\times$ allowing compiler-version drift)
relative to published \armfour{} reference~C, despite lacking
64-bit multiply, DSP, and SIMD instructions.  A second physical
RP2040 agrees to within 0.075\% on post-quantum operations, below
the shift a compiler-version change alone produces.
All code and data are released as an open-source benchmark suite.

\end{abstract}


\section{Introduction}
\label{sec:intro}

Classical public-key algorithms---RSA and elliptic-curve
cryptography (ECC)---will be rendered insecure by cryptographically
relevant quantum computers~\cite{shor1997}.
The ``harvest now, decrypt later'' threat model means that data
transmitted today must already be protected against future quantum
adversaries, a concern amplified by the billions of IoT devices
with operational lifespans of 10--20\,years deployed in critical
infrastructure such as power grids, medical systems, and industrial
control networks~\cite{liu2024postquantumcryptographyinternetthings}.

In August 2024, the U.S.\ National Institute of Standards and
Technology (NIST) published three post-quantum cryptography (PQC)
standards: FIPS~203 (ML-KEM, a lattice-based key-encapsulation
mechanism)~\cite{fips203}, FIPS~204 (ML-DSA, a lattice-based digital
signature algorithm)~\cite{fips204}, and FIPS~205 (SLH-DSA, a
hash-based signature scheme)~\cite{fips205}.
NIST subsequently published FIPS~206
(FN-DSA, a lattice-based signature scheme derived from FALCON) in
2025; we leave its evaluation on M0+ to future work.
NIST mandates that U.S.\ federal systems migrate to these
standards by 2035, and the EU Cyber Resilience Act similarly requires
security updates for IoT devices sold in Europe.

The ARM \armfour{} processor---the target of the widely cited pqm4
benchmarking framework~\cite{kannwischer2019pqm4,
kannwischer2024pqm4}---is well characterised for PQC workloads.
However, the even more constrained ARM \armmo{} powers a vast
population of low-cost IoT endpoints---smart meters, medical
sensors, and industrial field nodes---built around
microcontrollers whose unit price in volume is typically under
\$5 (the RP2040 die itself is around \$1; the Raspberry~Pi~Pico
board used here retails at about \$4).  We are explicit about where the RP2040 sits in the
RFC~7228 taxonomy~\cite{rfc7228}, which defines Class~0 as
$\ll$10\,KiB RAM and $\ll$100\,KiB Flash, Class~1 as
${\sim}$10\,KiB RAM and ${\sim}$100\,KiB Flash, and Class~2 as
${\sim}$50\,KiB RAM and ${\sim}$250\,KiB Flash.
With 264\,KB SRAM and 2\,MB flash the RP2040 is \emph{beyond}
Class~2 in memory; RFC~7228 deliberately defines no class above
Class~2.  It is constrained in the dimension this paper actually
measures---\emph{processing}: an ARMv6-M core with no 64-bit
multiply, no DSP or SIMD, and eight directly addressable
registers.  We therefore use the RP2040 as an \emph{instruction-set}
proxy for the ARMv6-M class rather than as a memory-constrained
device, and we report peak stack precisely so that the
memory question can be answered separately for genuinely
Class-1 and Class-2 parts (Section~\ref{sec:discussion}).  One
qualifier applies to every feasibility statement in this paper,
including those in Table~\ref{tab:recommendation}: our stack and
timing figures are for PQClean reference~C, not for
stack-reduced or assembly-optimised implementations.  Optimised
variants such as \texttt{mlkem-native} report substantially
lower peak stack, so our numbers should be read as an upper
bound on the memory cost of deployment, and a lower bound on
which device classes can host these algorithms.
The \armmo{} implements the ARMv6-M instruction set with
approximately 52 instructions, no 64-bit multiply
(\texttt{UMULL}/\texttt{SMULL}), no DSP or SIMD extensions, and
only eight directly accessible low registers (R0--R7).
Despite five prior studies exploring PQC on Cortex-M0 or M0+
hardware~\cite{halak2024, karmakar2018saber, bos2021masking,
li2023indocrypt, bos2022dilithium}, and despite broad surveys of
PQC for IoT~\cite{kumar2022securing} and of PQC support in
cryptographic libraries~\cite{ahmed2025library}, none benchmarked
the \emph{finalised} FIPS~203 and~204 standards; all used
pre-standardisation algorithm versions (Kyber, Dilithium, or Saber).
A claim of this kind cannot be established exhaustively, so we state
it as ``to the best of our knowledge'' throughout: we are aware of no
prior isolated algorithm-level benchmark of the finalised FIPS~203
and~204 standards on this core, and we would welcome correction.
This paper focuses on the two lattice-based standards (ML-KEM and
ML-DSA); the hash-based SLH-DSA (FIPS~205) has fundamentally
different performance characteristics---large signatures and
deterministic but slow signing---warranting a separate study.

\textbf{Contributions.} This paper makes five contributions,
spanning three measurement campaigns on two physical devices:
\begin{enumerate}
  \item \textbf{Timing and memory characterisation.}  To the best of
    our knowledge, the first isolated algorithm-level benchmarks of
    NIST-standardised \mlkem{} (FIPS~203) and \mldsa{} (FIPS~204) on
    \armmo{}, covering all six variants across key generation,
    encapsulation/signing, and decapsulation/verification, together
    with peak stack and flash footprint---a memory compatibility map
    for the PQClean reference implementations that determines
    feasibility for Class-1 IoT.  Energy is estimated, not measured
    (Table~\ref{tab:howtoread}).
  \item \textbf{The signing tail.}  A variance analysis of \mldsa{}
    signing latency quantifying the effect of FIPS~204 rejection
    sampling on constrained hardware, resolved to the $p_{99.9}$
    quantile over \num{10000} signatures, and showing that the
    $n{=}100$ tail estimate routinely used in embedded benchmarking
    understates the \num{10000}-signature $p_{99}$ by
    \SI{16}{\percent}, and its observed maximum by a factor of
    $2.1\times$.
  \item \textbf{Iteration recovery from wall-clock time.}  A technique
    that recovers the rejection-sampling iteration count of
    \emph{each individual signature} using only an uninstrumented
    microsecond wall clock---no cycle counter, no debugger, and no
    modification to the implementation under test---and separates
    fixed setup cost from marginal per-iteration cost.  We show that
    the natural least-squares objective is \emph{degenerate}, so the
    lattice is identified only with an additional criterion drawn
    from the algorithm's own structure, and we validate the
    underlying model against an instrumented rejection counter on a
    host build (Section~\ref{sec:latticegen}).  We delimit where the method
    works and where it must abstain.
  \item \textbf{What is build-invariant, and what is not.}  An
    independent replication on a separately written harness,
    toolchain, and optimisation level, combined with an inter-unit
    equivalence test on a second physical die.  Together these
    separate build sensitivity from manufacturing spread and show
    that a compiler-version change at fixed optimisation level moves
    results roughly \num{26} times more than the difference between
    two dies, with a change of optimisation level larger still
    (Section~\ref{sec:interunit})---so build configuration must be
    reported alongside any absolute figure.
  \item \textbf{Reproducible artefact.}  An open-source benchmark
    suite---firmware, serial capture scripts, raw data, and the
    analysis scripts that regenerate every table and figure in this
    paper---enabling the community to reproduce or extend these
    results on RP2040 hardware.
\end{enumerate}

\textbf{Scope.}  The cryptographic algorithms are the
\emph{workload} of this study, not its object.  We make no claim
about the security, parameter choices, or cryptanalysis of ML-KEM
or ML-DSA, and we propose no implementation improvement to either.
The object of study is the platform: how an ARMv6-M core lacking a
64-bit multiplier, DSP extensions, and SIMD absorbs a lattice
workload; how much of its SRAM and flash budget that consumes;
where the resulting latency distribution places a real-time
deadline; and how far those figures move under changes of
toolchain, optimisation level, and physical die.  The standardised
algorithms suit this purpose because they are fixed, widely
deployed, and now unavoidable in embedded firmware.

\begin{table}[t]
\centering
\caption{How to read this paper: the four standing caveats that
qualify every number reported here.  They are stated once, in
full, and referenced rather than repeated.}
\label{tab:howtoread}
\footnotesize
\begin{tabularx}{\columnwidth}{@{}lX@{}}
\toprule
\textbf{Caveat} & \textbf{Statement} \\
\midrule
Energy &
Not measured.  All energy figures are a unit conversion of
latency under a constant-power datasheet model
(\SI{0.0792}{\milli\watt\per\micro\second}); they carry no
information beyond the timing tables and are confined to
Appendix~\ref{sec:appendix-energy}. \\
\addlinespace
Build
configuration &
Absolute figures are properties of a build, not of an algorithm.
A compiler-version change at fixed \texttt{-O3} moves latency by
\SIrange{1.4}{2.5}{\percent} and an optimisation-level change by
\SIrange{4.3}{15.8}{\percent}, both larger than the inter-unit
effect (\SI{0.075}{\percent}).  Toolchain and flags are reported
with every figure. \\
\addlinespace
Reference
implementations &
We measure PQClean reference~C, not optimised assembly.
Figures are upper bounds on latency and memory; optimised
implementations move them down, never up, so feasibility
verdicts are conservative. \\
\addlinespace
Tail
quantiles &
Quantiles from $n{=}100$ are lower bounds.  The rejection loop
is geometric and has no finite worst case; deadline analysis
should use the model, not a measured maximum
(Section~\ref{sec:results}). \\
\bottomrule
\end{tabularx}
\end{table}

Table~\ref{tab:howtoread} states the four caveats that qualify
every number in this paper.  Rather than restate them at each
point of use, we give them once here and refer back.

The remainder of the paper is structured as follows.
Section~\ref{sec:background} provides background on ML-KEM, ML-DSA,
and the \armmo{} architecture.
Section~\ref{sec:related} surveys related work.
Sections~\ref{sec:methodology} and~\ref{sec:results} describe the
experimental methodology and present results.
Section~\ref{sec:discussion} discusses practical implications, and
Section~\ref{sec:conclusion} concludes.

\section{Background}
\label{sec:background}

\subsection{ML-KEM (FIPS~203)}
ML-KEM is a lattice-based key-encapsulation mechanism derived from
CRYSTALS-Kyber.  It provides IND-CCA2 security through the
module-learning-with-errors (Module-LWE) problem.  FIPS~203 defines
three parameter sets---ML-KEM-512, 768, and 1024---offering
roughly 128, 192, and 256 bits of classical security,
respectively~\cite{fips203}.  Public keys range from 800 to
\num{1568}\,bytes, and ciphertexts from 768 to
\num{1568}\,bytes.  All variants produce a 32-byte shared
secret.  The core computational kernel is the Number Theoretic
Transform (NTT) over the polynomial ring $\mathbb{Z}_q[x]/(x^{256}+1)$
with $q = 3329$.

\subsection{ML-DSA (FIPS~204)}
ML-DSA is a lattice-based digital signature scheme derived from
CRYSTALS-Dilithium, providing EUF-CMA security via the
Module-LWE and Module-SIS problems.  FIPS~204 defines three
parameter sets---ML-DSA-44, 65, and 87---at security levels
roughly equivalent to AES-128, AES-192, and AES-256,
respectively~\cite{fips204}.
A critical characteristic of ML-DSA is its use of the
\emph{Fiat--Shamir with Aborts} paradigm: signing involves rejection
sampling, where the signer generates a candidate signature and
discards it if a norm check fails.  The expected number of iterations
is 4.25, 5.1, and 3.85 for levels 44, 65, and 87,
respectively~\cite{fips204, dilithium2022}.
Note that this ordering is \emph{non-monotonic} in the security
level: level~65 requires more iterations on average than
level~87, because the expected count depends on the
$(\gamma_1,\gamma_2,\beta)$ triple rather than on the matrix
dimension.  The mechanism is a trade-off between $\beta$ and
$\gamma_2$: a signature is retried when any of the $256k$
low-order coefficients falls within $\beta$ of the $\gamma_2$
boundary, so the per-attempt rejection probability grows with
the matrix dimension $k$ but shrinks with the ratio
$\beta/\gamma_2$.  From level~65 to level~87, $k$ grows from 6
to 8, which alone would mean more rejections; but $\beta$ falls
from 196 to 120 at unchanged $\gamma_2$, cutting
$\beta/\gamma_2$ from $7.5\times10^{-4}$ to
$4.6\times10^{-4}$.  The second effect dominates, so
$\mathbb{E}[k]$ drops from 5.10 to 3.85.  Level~65 is the worst
case because it pairs the largest $\beta$ with a mid-sized
matrix.
This makes signing time inherently non-deterministic---a key concern
for real-time IoT applications.

\subsection{ARM Cortex-M0+ and the RP2040}
The ARM \armmo{} implements the ARMv6-M architecture, the most
constrained 32-bit ARM profile.  Compared to the ARMv7E-M
(\armfour{}) used in the pqm4 framework, the \armmo{} lacks three
capabilities critical for PQC performance:
\begin{itemize}
  \item \emph{No 64-bit multiply result, and no multiply--accumulate:} the
  \texttt{UMULL} and \texttt{SMULL} instructions that produce a 64-bit product
  from two 32-bit operands are absent, as is \texttt{MLA}/\texttt{MLS} --- the
  core offers only \texttt{MULS}, a $32\times32\rightarrow32$ multiply, so every
  accumulation costs a separate instruction. Which of these absences binds
  depends on the modulus width, and the two standards fall on opposite sides of
  that division. ML-DSA works over $q=8\,380\,417$ with 32-bit coefficients, and
  the reference Montgomery reduction consumes a 64-bit intermediate product;
  on ARMv6-M that product must be assembled from several
  $32\times32\rightarrow32$ multiplies and adds, so here the missing long
  multiply is the operative penalty. ML-KEM works over $q=3329$ with 16-bit
  coefficients, where the product of two coefficients already fits in 32 bits;
  the 64-bit multiply is therefore not on the critical path, and the M4's
  advantage in the ML-KEM reduction steps comes from the DSP extension named
  below rather than from \texttt{SMULL}.
  \item \emph{No DSP/SIMD, which is what costs ML-KEM its reduction throughput:}
  the Cortex-M4 packs two 16-bit coefficients per register and operates on both
  at once, using the halfword multiply--accumulate family
  (\texttt{SMULBB}, \texttt{SMLABB}) and its dual-MAC variants
  (\texttt{SMUAD}, \texttt{SMLAD}, \texttt{SMLALD}) together with packed
  halfword arithmetic (\texttt{SADD16}, \texttt{SSUB16}, \texttt{PKHBT}). This
  is the instruction family that optimised Cortex-M4 NTT implementations rely on
  for both the butterfly and the Montgomery and Barrett reduction steps
  \cite{alkim2020cortexm4,abdulrahman2024fastclean,abdulrahman2022multimoduli}.
  The Cortex-M0+ has none of it, so each 16-bit coefficient is reduced
  individually in the general-purpose datapath --- the reason the reduction cost
  in ML-KEM scales with coefficient count on this core rather than with
  half of it \cite{li2023indocrypt,chung2021ntt}.
  \item \emph{Limited register file:} only eight low registers (R0--R7) are
  directly accessible with the Thumb ISA, versus the full 16-register set on
  Cortex-M4. No barrel shifter is available, so shifts that the M4 folds into a
  data-processing operand cost a separate instruction here.
\end{itemize}

The RP2040 (Raspberry Pi Pico) integrates a dual-core \armmo{}
at a datasheet-maximum 133\,MHz (default \texttt{clk\_sys}
\SI{125}{\mega\hertz}) with 264\,KB\,SRAM and 2\,MB QSPI flash.
Critically, the RP2040 implements the \emph{single-cycle}
$32 \times 32 \rightarrow 32$-bit multiplier option---many
\armmo{} implementations use a 32-cycle multiplier instead.
This design choice is what makes PQC computationally feasible
on this platform~\cite{rp2040datasheet}; Section~\ref{sec:multiplier}
estimates what the 32-cycle alternative would cost.

\section{Related Work}
\label{sec:related}

\subsection{PQC on Cortex-M4}
The pqm4 project~\cite{kannwischer2019pqm4, kannwischer2024pqm4}
is the de facto benchmarking standard for PQC on ARM
microcontrollers, providing cycle counts for ML-KEM, ML-DSA,
SLH-DSA, and additional NIST candidates on \armfour{} targets;
its published \texttt{benchmarks.csv} figures, which we compare
against in Section~\ref{sec:results}, are measured on a
NUCLEO-L4R5ZI at \SI{24}{\mega\hertz}.
Subsequent work has extended M4 benchmarks to
\armfour{}/M7 porting via the SLOTHY
optimiser~\cite{abdulrahman2025slothy}, efficient sparse
polynomial multiplication for ML-DSA~\cite{zheng2024espmd},
and ML-DSA on 16-bit MSP430 processors~\cite{shin2025msp430}.
The pqmx framework~\cite{abdulrahman2024fastclean} targets Cortex-M55/M85
with Helium MVE extensions---skipping the \armmo{} entirely.
No ``pqm0+'' equivalent exists.

\subsection{Prior M0+/M0 Work}
Five prior works have explored PQC on Cortex-M0 or M0+ hardware.
We summarise each and explain the gap our work fills.

\textbf{Halak et al.}~\cite{halak2024} benchmarked
pre-standardisation Kyber-512 and Dilithium-2 on the Raspberry Pi
Pico~W within TLS handshakes using MbedTLS with experimental liboqs
integration (IEEE Access, 2024).
They measured TLS-level latency, energy, and computation costs but
did not isolate algorithm-level cycle counts, did not test all
security levels, did not profile stack or flash usage, and used
pre-standardisation algorithm versions.

\textbf{Karmakar et al.}~\cite{karmakar2018saber} implemented
Saber KEM on a Cortex-M0 with a memory-efficient design achieving
4.8--7.5 million cycles (TCHES, 2018).
This proved lattice-based KEMs can fit on M0-class devices, but
Saber was not selected by NIST for standardisation.

\textbf{Bos, Gourjon, Renes et al.}~\cite{bos2021masking}
implemented the first complete masked (side-channel-protected) Kyber
decapsulation on an NXP FRDM-KL82Z (\armmo{}) and validated with
\num{100000} power traces (TCHES, 2021).
Their focus was on side-channel protection overhead, not raw
performance benchmarking, and they used pre-standardisation Kyber
without ML-DSA.

\textbf{Li, Wang \& Wang}~\cite{li2023indocrypt} optimised
NTT-based polynomial multiplication for Kyber and Saber on
Cortex-M0/M0+ using hybrid k-reduction and multi-moduli NTTs,
achieving a ${\sim}2.9\times$ speedup for Saber's polynomial
multiplication (INDOCRYPT, 2023), building on the foundational
NTT work by Chung et al.~\cite{chung2021ntt} for M0/M0+.
They benchmarked only the polynomial multiplication
component, not complete KEM or signature operations.

\textbf{Bos, Renes \& Sprenkels}~\cite{bos2022dilithium} created
a compact Dilithium implementation requiring $<$7\,KB for signing
and $<$3\,KB for verification, targeting M0+-class memory
constraints (AFRICACRYPT, 2022).
Despite targeting M0+ memory profiles, they benchmarked only on
\armfour{} (pqm4) and used pre-standardisation Dilithium.

\textbf{Concurrent work on benchmarking methodology.}
The five studies above are the prior \armmo{}/M0 implementation
literature.  The two that follow are a separate category: they
were developed concurrently with this work and concern how PQC
measurements should be reported rather than M0+ implementations,
so they are not part of that count.

\textbf{Benchmarking methodology for ML-DSA.}
Developed concurrently with this work and since published at
MAgiCS~2026, Riou et al.~\cite{pqshield2026pitfalls}
catalogue pitfalls that make published ML-DSA measurements
non-comparable, and propose a reporting standard based on
worst-case execution time rather than mean latency.  Their
measurements are taken on a Cortex-M33 with an instrumented
build that exposes the internal rejection counter.  Our
measurement is complementary in both platform and method: we
recover the same rejection structure on ARMv6-M from an
uninstrumented \SI{1}{\micro\second} wall clock, and we adopt
their disclosure requirements (interface, message length,
hedged-versus-deterministic variant, and percentile reporting)
throughout Section~\ref{sec:methodology}.
Turino et al.~\cite{turino2025pqcleo} likewise argue for
standardised evaluation of PQC on IoT networks, but at the
network rather than the primitive level.
Table~\ref{tab:concurrent} places both alongside this work.

\begin{table}[t]
\centering
\caption{Concurrent methodology work on PQC evaluation.  These
  papers propose how PQC should be measured and reported, rather
  than reporting implementation figures, so they are compared
  separately from the implementation studies of
  Table~\ref{tab:priorwork}.}
\label{tab:concurrent}
\footnotesize
\setlength{\tabcolsep}{3pt}
\newcolumntype{Y}{>{\raggedright\arraybackslash}X}
\begin{tabularx}{\columnwidth}{@{}lYYY@{}}
\toprule
 & \textbf{Riou et al.}~\cite{pqshield2026pitfalls} &
   \textbf{Turino et al.}~\cite{turino2025pqcleo} &
   \textbf{This work} \\
\midrule
Level of
analysis
& Primitive
& Network / protocol
& Primitive \\
\addlinespace
Platform
& Cortex-M33
& IoT networks (platform-agnostic)
& Cortex-M0+ (ARMv6-M) \\
\addlinespace
Rejection
counts
& Instrumented build exposing the internal counter
& Not addressed
& Recovered from an uninstrumented \SI{1}{\micro\second} clock \\
\addlinespace
Reporting
proposal
& Worst-case execution time over mean latency
& Standardised network-level evaluation
& Adopts~\cite{pqshield2026pitfalls}; adds percentile and
  provenance disclosure \\
\bottomrule
\end{tabularx}
\end{table}

\textbf{Summary.} All five prior M0+/M0 studies used
pre-standardisation algorithm versions (Kyber, Dilithium, or Saber).
None provide isolated, algorithm-level benchmarks of the finalised
FIPS~203 and~204 standards with stack, flash, and energy metrics
across all security levels.
Table~\ref{tab:priorwork} summarises the key differences.

\begin{table*}[t]
\centering
\caption{Comparison with prior M0+/M0 PQC work.  ``Std.''\ indicates
  whether the finalised FIPS~203/204 standards were benchmarked.}
\label{tab:priorwork}
\footnotesize
\setlength{\tabcolsep}{4pt}
\begin{tabular}{lcccccc}
\toprule
\textbf{Work} & \textbf{Year} & \textbf{Hardware} &
  \textbf{Algorithm} & \textbf{Std.} &
  \textbf{Isolated metrics} & \textbf{All levels} \\
\midrule
Halak et al.~\cite{halak2024}
  & 2024 & RP2040 (M0+) & Kyber/Dilithium & No & No (TLS-level) & No \\
Karmakar et al.~\cite{karmakar2018saber}
  & 2018 & Cortex-M0 & Saber & No & Yes & N/A \\
Bos et al.~\cite{bos2021masking}
  & 2021 & NXP KL82Z (M0+) & Masked Kyber & No & Yes & No \\
Li et al.~\cite{li2023indocrypt}
  & 2023 & Cortex-M0/M0+ & Kyber/Saber poly.\ mul. & No & Partial & No \\
Bos et al.~\cite{bos2022dilithium}
  & 2022 & Cortex-M4 (M0+ target) & Dilithium & No & Yes & No \\
\midrule
\textbf{This work}
  & \textbf{2026} & \textbf{RP2040 (M0+)} & \textbf{ML-KEM / ML-DSA}
  & \textbf{Yes} & \textbf{Yes} & \textbf{Yes} \\
\bottomrule
\multicolumn{7}{l}{\scriptsize ``All levels'' records coverage of the three
  FIPS~203/204 parameter sets.  ``N/A'' for Karmakar et al.\ indicates that
  Saber was never}\\
\multicolumn{7}{l}{\scriptsize standardised, so coverage of a standardised
  level structure does not apply to that work.}\\
\end{tabular}
\end{table*}

\subsection{Broader PQC Landscape}
Several recent works study PQC on IoT but do not target the M0+
processor class.
Lopez et al.~\cite{lopez2025qce} evaluated PQC on Raspberry Pi
3B+ and~5 (Cortex-A, 64-bit Linux)---a fundamentally different
class from bare-metal M0+.
Grassl et al.~\cite{grassl2024iotbds} benchmarked PQC on
Raspberry Pi models~1B through~4B, all running Cortex-A
processors.
Banegas et al.~\cite{banegas2021ota} studied PQC for
over-the-air firmware updates on Cortex-M class devices but
provided no M0+-specific cycle counts.
Liu, Ramachandran, and Jurdak~\cite{liu2024postquantumcryptographyinternetthings} provide a comprehensive
survey of PQC for IoT with 86 citations but identify M0+-class
benchmarking as an open problem.
The Schwabe MSR talk~\cite{schwabe2019msr} explicitly identified
Cortex-M0 as a research target in 2019; as of 2026, this gap
remains unfilled for standardised algorithms.
On the commercial side, PQShield \emph{advertises}
PQMicroLib-Core~\cite{pqshield2026} as achieving ML-KEM in
\SI{5}{\kilo\byte} of RAM on Cortex-M.  We record this as a
vendor claim rather than evidence: it is a product announcement
with no accompanying technical report, no statement of which
parameter set, interface, or measurement method the figure
refers to, and no independent replication.  We cite it only to
indicate that commercial effort is directed at exactly the
memory pressure our reference~C figures exhibit, and no claim in
this paper rests on it.
Dinu~\cite{dinu2025migration} analysed the migration path from
ECDSA to ML-DSA, providing additional context for the classical
versus post-quantum comparison in our results.

\section{Experimental Methodology}
\label{sec:methodology}

\subsection{Hardware Platform}
\label{sec:hardware}
All experiments were conducted on a Raspberry Pi Pico~H
(RP2040): dual-core ARM~\armmo{}, 264\,KB SRAM,
2\,MB QSPI flash, with the single-cycle $32 \times 32 \rightarrow
32$-bit multiplier~\cite{rp2040datasheet}.  Only one core was
used for benchmarking; the second core remained idle.
The core clock was left at the Pico~SDK default
\texttt{clk\_sys}${=}\SI{125}{\mega\hertz}$
(the firmware never calls \texttt{set\_sys\_clock\_khz()});
133\,MHz is the RP2040 datasheet \emph{maximum}, not the
configured frequency.  The value is read on-device via
\texttt{clock\_get\_hz(clk\_sys)} and emitted in the capture
log header.  All cycle counts in this paper are derived at
\SI{125}{\mega\hertz}.

\subsection{Software Stack}
\textbf{PQC implementations.}
We used the PQClean~\cite{pqclean} reference~C implementations
of \mlkem{} and \mldsa{}, compiled with \texttt{-mcpu=cortex-m0plus
-mthumb} using \texttt{arm-none-eabi-gcc}~12.2.1 at
\texttt{-O3}, the \texttt{Release} default inherited from the
Pico~SDK.  The replication campaign of
Section~\ref{sec:replication} pins \texttt{-Os} explicitly
across every translation unit, which allows the effect of the
optimisation level to be read off directly.
PQClean provides portable, well-tested, pure~C reference code
without platform-specific assembly optimisations, forming the
basis of the pqm4 framework's reference implementations.
We deliberately chose reference~C to establish a \emph{baseline}
for \armmo{} capability; optimised assembly implementations would
improve these numbers but are beyond the scope of this study.
We note that the PQClean repository is transitioning to read-only
status (scheduled July 2026) as the community migrates to the
PQCA \texttt{mlkem-native} and \texttt{mldsa-native}
packages~\cite{pqca2026}.  Our results were obtained from PQClean at commit
\texttt{3730b32a} (pinned prior to this transition); the
reference~C implementations are algorithmically identical to the
PQCA successors (\texttt{mlkem-native} and \texttt{mldsa-native}),
which derive from the same NIST-submitted code.  The exact commit
hash and build instructions are provided in the reproducibility
repository to ensure long-term replicability.

The RP2040's ring oscillator (ROSC) was used for random number
generation via a thin \texttt{randombytes\_pico.c} adapter.
While the ROSC is adequate for benchmarking purposes---it does
not affect execution timing since \mlkem{} and \mldsa{} call
\texttt{randombytes()} a fixed number of times per
operation---it is not cryptographically certified.  Production
deployments would require a DRBG seeded from a certified
entropy source (e.g., NIST SP~800-90A).

\textbf{Classical baselines.}
RSA-2048, ECDSA\,P-256, and ECDH\,P-256 were benchmarked
using mbedTLS~3.6.0, bundled with the Raspberry~Pi Pico~SDK~v2.2.0
as the \texttt{pico\_mbedtls} library.  Our custom
\texttt{mbedtls\_config.h} (included in the repository) enables
only the required algorithms---RSA PKCS\#1~v1.5, ECDSA, ECDH on
secp256r1---with no hardware-accelerated big-number operations,
providing a direct comparison on identical hardware.
Note that PQClean does not include classical algorithms; the
classical baselines are entirely from mbedTLS.

\subsection{Porting Path and Build Configuration}
PQClean's reference~C requires no algorithmic modification to
build for ARMv6-M; the integration cost is entirely in the
platform interface.  Three changes were needed.
First, \texttt{randombytes()}: PQClean ships a POSIX
implementation reading \texttt{/dev/urandom}, which we replaced
with \texttt{randombytes\_pico.c}, a 40-line adapter over the
RP2040 ring oscillator's \texttt{randombit} register.
Second, the linker script: the Pico~SDK default reserves a
2\,KB stack per core, far below \mldsa{}-87's requirement, so
we supply \texttt{memmap\_custom.ld} raising the main stack to
192\,KB and placing \texttt{\_\_StackLimit} at a known symbol
for the stack-painting measurement.
Third, \texttt{stdio}: results are emitted over USB-CDC, which
requires a \SI{10}{\milli\second} yield after each line so the
CDC endpoint does not overflow during long captures; this
\texttt{sleep\_ms()} lies outside every timed region.
No source file under \texttt{crypto\_kem/} or
\texttt{crypto\_sign/} was edited.  The build uses
\texttt{arm-none-eabi-gcc}~12.2.1 with
\texttt{-mcpu=cortex-m0plus -mthumb} at \texttt{-O3}; the
complete \texttt{CMakeLists.txt}, linker script, and adapter
are in the reproducibility repository.

\subsection{Timing Methodology}
Execution time was measured using the Pico~SDK's
\texttt{time\_us\_32()} function, which provides
\SI{1}{\micro\second} resolution from the RP2040's hardware
timer.  Each deterministic operation (key generation,
encapsulation, decapsulation, verification) was executed 30
times; ML-DSA signing was executed 100 times per security
level to capture rejection-sampling variance.
No explicit warm-up iterations were performed.
The Cortex-M0+ core itself lacks instruction and data caches,
branch prediction, and speculative execution; the only caching
layer is the RP2040's 16\,KB two-way set-associative XIP flash
cache, which reaches steady state within the first few function
calls.  The consistently low CV ($<$1.5\% for deterministic
operations) confirms that first-iteration cache-cold effects are
negligible across our 30-run samples.
The timed region contains the cryptographic call and nothing
else: \texttt{time\_us\_32()} is read immediately before and
immediately after the single \texttt{PQCLEAN\_*} invocation,
inside the same C block.  Buffer allocation, key generation
preceding a signing measurement, the \texttt{memcmp}/verify
correctness check, CSV formatting, and the USB yield all occur
outside the timestamps.  Timer read overhead was measured at
\SI{2}{\micro\second} for a back-to-back pair, i.e.\
$0.02\%$ of the shortest operation reported here
(\mlkem{}-512 key generation, \SI{9.94}{\milli\second}), and is not
subtracted.
Following the disclosure requirements of
Riou et al.~\cite{pqshield2026pitfalls}, we state the
signing configuration explicitly: we measure the \emph{Pure}
\mldsa{} interface (\texttt{crypto\_sign\_signature}, empty
context string), in the \emph{hedged} variant --- PQClean's
\texttt{crypto\_sign\_signature\_ctx} calls
\texttt{randombytes()} for the 32-byte $\rho''$ seed
unconditionally, so every measurement includes fresh
randomness generation --- over a fixed \num{128}-byte message
(\texttt{test\_message} in \texttt{src/bench\_mldsa.c}).
The message digest $\mu$ absorbs
$t\!r \,\Vert\, M'$, where $M' =
\mathtt{0x00} \,\Vert\, \mathtt{0x00} \,\Vert\, M$ for an empty
context string; at \SI{128}{\byte} this is \num{194} bytes, which
exceeds the \SI{136}{\byte} SHAKE256 rate and therefore spans two
Keccak-\textit{f} permutations rather than one.  A message of
\SI{70}{\byte} or fewer would fit a single block and reduce
absolute signing times by roughly one permutation, without
affecting the variance structure that is the subject of
Section~\ref{sec:results}.
All measurements were performed on real hardware---no simulators
or emulators were used.
Correctness is verified at runtime: for \mlkem{}, the encapsulated
and decapsulated shared secrets are compared via \texttt{memcmp};
for \mldsa{}, every generated signature is verified against the
public key before the next iteration.  All operations passed
these checks across every run.  Additionally, the PQClean
implementations we use pass PQClean's own test harness (which
includes NIST KAT vector comparisons) on the build host prior
to cross-compilation.
Because \texttt{Release} implies \texttt{-DNDEBUG}, we audited
every translation unit in the firmware link for
\texttt{assert()}, \texttt{static\_assert}, and
\texttt{<assert.h>}: there are none.  The sole PQClean source at
the pinned commit that uses \texttt{assert()} is
\texttt{common/randombytes.c}, which targets desktop operating
systems and is not compiled here, being replaced by our
\texttt{randombytes\_pico.c}.  The runtime checks described
above are unconditional \texttt{memcmp} and verification calls
rather than assertions, so no correctness check is removed by
the optimised build.

\subsection{Memory Profiling}
Peak stack usage was measured using the stack-painting technique:
the region between the linker symbol \texttt{\_\_StackLimit}
and the current stack pointer is initialised with a sentinel
value (\texttt{0xDEADBEEF}) before each operation, and the
high-water mark is determined by scanning upward from
\texttt{\_\_StackLimit} for the first non-sentinel word after
execution.  Because painting begins at the SP inside the
benchmark frame, the reported figure is the stack consumed by
the cryptographic call itself and excludes the harness frame,
the interrupt stack, and the USB/stdio machinery.
The sentinel word at \texttt{\_\_StackLimit} is checked after
each measurement; a non-sentinel value flags a painted-region
overflow and invalidates the reading.  This guard is essential:
without it a stack requirement exceeding the painted region
saturates silently, and the resulting figure is a lower bound
reported as a measurement.
Key material, ciphertexts, and signatures are caller-allocated
in the benchmark frame and are therefore \emph{not} counted in
the stack column; they are added separately in the Total~RAM
column of Table~\ref{tab:memory}.
Flash footprint (text, data, BSS) was obtained using
\texttt{arm-none-eabi-size -A} on the compiled ELF binary.

\subsection{Cycle Counts}
The RP2040's Cortex-M0+ core lacks instruction and data caches,
branch prediction, and speculative execution; all instructions
execute in deterministic time relative to the system clock.
We therefore derive cycle counts as
$\text{cycles} = t\,(\mu\text{s}) \times 125$, where 125 is the
\emph{configured} core frequency in MHz
(Section~\ref{sec:methodology}).
Wall-clock times are unaffected by this choice:
\texttt{time\_us\_32()} is driven by the \SI{1}{\mega\hertz}
watchdog tick, not by \texttt{clk\_sys}.
Tables~\ref{tab:mlkem}, \ref{tab:mldsa},
and~\ref{tab:classical} report both wall-clock time and kilocycles
(kc) to facilitate direct comparison with the pqm4 benchmarking
framework~\cite{kannwischer2024pqm4}, which reports Cortex-M4 cycle
counts.  Tables~\ref{tab:replication} and~\ref{tab:tail} concern the
replication campaign and its signing tail and are reported in
milliseconds only: campaign~B was built at \texttt{-Os}, so its
cycle counts are not comparable with the \texttt{-O3} figures
elsewhere in this section.  Two measurement campaigns are
referred to throughout: \emph{campaign~A}, the primary
\texttt{-O3} campaign whose results are reported in this
section, and \emph{campaign~B}, an independent \texttt{-Os}
replication with larger samples, described in full in
Section~\ref{sec:replication}.

\subsection{Energy Estimation}
Energy per operation was estimated using the RP2040 datasheet
power model: \SI{3.3}{\volt} supply, \SI{24}{\milli\ampere}
typical active-mode current, yielding
\SI{79.2}{\milli\watt} active power~\cite{rp2040datasheet}.
Energy is calculated as
$E\,(\mu\text{J}) = t\,(\mu\text{s}) \times 0.0792$.
This constant-power model represents a conservative upper bound:
the RP2040 draws less current during memory stalls and idle
wait states than during sustained ALU computation, so actual
energy consumption is likely lower than reported.
We note that Halak et al.~\cite{halak2024} employed the same
datasheet-based estimation methodology for the RP2040 in their
IEEE Access study.
Because the power draw is assumed constant, energy is strictly
proportional to execution time: the energy figures of
Appendix~\ref{sec:appendix-energy} are a re-scaling of the timing
results by \SI{79.2}{\micro\joule\per\milli\second} and are
reported for convenience rather than as an independent
measurement.  The
\emph{relative} rankings between algorithms remain valid
regardless of the power model's absolute accuracy, since all
algorithms were measured on identical hardware at the same clock
speed.

\subsection{Statistical Measures}
For each operation we report: mean, minimum, maximum, standard
deviation, and coefficient of variation (CV\,=\,SD/mean$\times$100).
For ML-DSA signing, we additionally report the 95th and 99th
percentiles to characterise worst-case latency for time-critical
applications.

\subsection{Data Provenance}
Three campaigns underlie this paper.  Campaign~A is the primary
campaign reported throughout ($n{=}30$ deterministic,
$n{=}100$ signing, \texttt{-O3}); campaign~B is an independent
replication on a separately written harness at \texttt{-Os} with
$n{=}\num{1000}/\num{10000}$ (Section~\ref{sec:replication});
campaign~C is an inter-unit comparison run on two physical devices
under one firmware build (Section~\ref{sec:interunit}).  Each is
reported separately and none is pooled with another: campaign~B's
value lies in having been written independently, and campaign~C's
in holding the build fixed while varying the device, so combining
them would destroy the property that makes each informative.

Within campaign~A, every headline figure comes from one
authoritative run conducted after the firmware build
configuration was frozen; we do not pool it with the earlier
exploratory runs.  Those preliminary runs are archived under
\texttt{results/old\_data/} and agree with the authoritative run
to within the expected run-to-run variance ($<2\%$ on
deterministic operations).  The residual differences---%
\mlkem{}-512 full handshake at \SI{36.3}{\milli\second}
preliminary against \SI{35.7}{\milli\second} final, for
instance---are attributable to a firmware rebuild with updated
toolchain settings, and are the same order as the build
sensitivity quantified in Sections~\ref{sec:replication}
and~\ref{sec:interunit} rather than a separate source of
uncertainty.  The \texttt{reproduce.py compare} script performs a
statistical comparison between any two runs.

Record-keeping tightened over the three campaigns, and campaign~C
reflects the current standard.  Its protocol was fixed before the
second unit was measured---sample sizes, the $\pm1\%$ equivalence
margin, the TOST procedure, and the minimum-$n$ requirement for a
difference verdict were all set in advance---and the analysis
script emits a verdict per operation without manual intervention.
Every capture records the device's hardware unique ID, its
\texttt{clk\_sys} frequency read at boot, the firmware build
stamp, and a reconciliation of rows received against rows
expected; captures with any shortfall were discarded and re-run.
One field we did not capture is the RP2040 die revision
(B0/B1/B2).  This was an oversight on our part, not a limitation
of the platform: the SDK exposes
\texttt{rp2040\_chip\_version()}, which returns~1 for B0/B1
and~2 for B2, and \texttt{rp2040\_rom\_version()}, which
distinguishes all three steppings---and our firmware simply did
not read either.  Because the boards have since been
re-flashed, we cannot recover the revision of the units used in
Campaigns~A and~B after the fact, and we decline to infer a
stepping from purchase date.  A replication seeking to reproduce
our absolute figures to better than the inter-unit bound of
Section~\ref{sec:interunit} should log both values at boot.  We
have added both calls to the provenance header in the released
firmware \emph{source}; the binaries used for the campaigns
reported here predate that change, so no measurement in this
paper carries a recorded stepping.
Campaigns~A and~B predate the unique-ID and reconciliation
tooling.  Both ran on the same physical unit, but that is
established by laboratory record rather than by an identifier
embedded in the capture, and the capture logs of campaign~A carry
a hard-coded platform banner rather than a measured clock; the
\SI{125}{\mega\hertz} operating point stated in
Section~\ref{sec:hardware} is the configured system clock, and
campaign~C confirms it by direct readback.

\section{Results}
\label{sec:results}

\subsection{ML-KEM Timing}
Table~\ref{tab:mlkem} presents timing results for all three
\mlkem{} security levels.  All operations exhibit low variance
(CV\,$<$\,1.5\%), confirming deterministic execution.
With $n=30$ runs and CV\,$<$\,1.5\%, the 95\% confidence interval
for each mean is within $\pm$0.60\% of the reported value
(worst case; CV\,=\,1.47\% for \mlkem{}-512 key generation).
This is the \num{4000}-resample percentile bootstrap, which makes
no normality assumption and is the widest of the three intervals we
computed.\footnote{For the same worst case the Student-$t$ interval
on 29 degrees of freedom is $\pm$0.55\% and the normal
approximation $\pm$0.53\%, against the bootstrap's $\pm$0.60\%.  We
report the widest because the timing samples are mildly
right-skewed, so the analytic intervals understate the upper tail;
the three agree to within 0.10 percentage points of the mean in
every case, so the choice affects no conclusion.}  All three are
derived from the observed sample spread, not from the timer's
resolution.  The RP2040 timer quantises at
\SI{1}{\micro\second}, which for a uniform quantisation error
contributes at most \SI{0.003}{\percent} of the mean to the standard
deviation of any operation reported here, which is at most
\SI{1.8}{\percent} of that operation's observed CV.  The
reported precision therefore reflects genuine run-to-run stability
of the device rather than a resolution artefact.

\begin{figure}[t]
\centering
\includegraphics[width=\columnwidth]{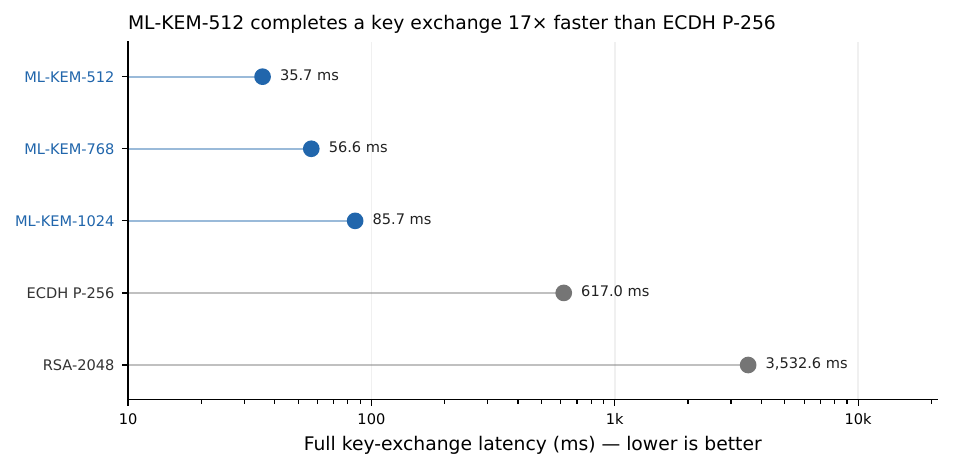}
\caption{Key exchange latency comparison on ARM Cortex-M0+.
  ML-KEM-512 achieves a full handshake (keygen + encaps + decaps)
  in \SI{35.7}{\milli\second}, $17\times$ faster than a complete
  ECDH\,P-256 key agreement (both parties' keygen and shared-secret
  computation).}
\label{fig:timing}
\end{figure}

ML-KEM-512 completes key generation in \SI{9.94}{\milli\second},
encapsulation in \SI{11.53}{\milli\second}, and decapsulation in
\SI{14.23}{\milli\second}, yielding a full key-exchange handshake
(keygen~+~encaps~+~decaps) of \SI{35.7}{\milli\second}.
Scaling to higher security levels, ML-KEM-768 requires
\SI{56.6}{\milli\second} and ML-KEM-1024 requires
\SI{85.7}{\milli\second} for a complete handshake.
Figure~\ref{fig:timing} shows these times against the classical
baselines of Table~\ref{tab:classical}.

\begin{table}[t]
\centering
\caption{ML-KEM timing on ARM Cortex-M0+ (RP2040, \SI{125}{\mega\hertz}).
  30 runs per operation.  Cycles = $t\,(\mu s) \times 125$.}
\label{tab:mlkem}
\footnotesize
\setlength{\tabcolsep}{3pt}
\begin{tabular}{llrrrrr}
\toprule
\textbf{Variant} & \textbf{Op.} & \textbf{Mean} &
  \textbf{Cycles} & \textbf{Min} & \textbf{Max} & \textbf{CV} \\
 & & \textbf{(ms)} & \textbf{(kc)} & \textbf{(ms)} & \textbf{(ms)} & \\
\midrule
\multirow{3}{*}{\shortstack[l]{ML-KEM\\-512}}
  & KeyGen  &  9.94 & \num{1243} &  9.89 & 10.68 & 1.47\% \\
  & Encaps  & 11.53 & \num{1441} & 11.47 & 11.60 & 0.17\% \\
  & Decaps  & 14.23 & \num{1779} & 14.12 & 14.28 & 0.19\% \\
\cmidrule{2-7}
  & \textit{Total} & \textit{35.71} & \textit{\num{4464}} & & & \\
\midrule
\multirow{3}{*}{\shortstack[l]{ML-KEM\\-768}}
  & KeyGen  & 16.02 & \num{2002} & 15.97 & 16.33 & 0.53\% \\
  & Encaps  & 18.58 & \num{2323} & 18.49 & 18.62 & 0.11\% \\
  & Decaps  & 22.02 & \num{2752} & 21.95 & 22.05 & 0.07\% \\
\cmidrule{2-7}
  & \textit{Total} & \textit{56.62} & \textit{\num{7077}} & & & \\
\midrule
\multirow{3}{*}{\shortstack[l]{ML-KEM\\-1024}}
  & KeyGen  & 25.18 & \num{3148} & 25.13 & 25.69 & 0.45\% \\
  & Encaps  & 28.10 & \num{3512} & 28.07 & 28.15 & 0.06\% \\
  & Decaps  & 32.45 & \num{4056} & 32.42 & 32.53 & 0.06\% \\
\cmidrule{2-7}
  & \textit{Total}\textsuperscript{\dag} & \textit{85.73} & \textit{\num{10716}} & & & \\
\bottomrule
\multicolumn{7}{l}{\scriptsize \textsuperscript{\dag}Total =
  keygen + encaps + decaps, measured sequentially on}\\
\multicolumn{7}{l}{\scriptsize \phantom{\textsuperscript{\dag}}%
  a single device (one party's full handshake contribution).}
\end{tabular}
\end{table}

\subsection{ML-DSA Timing and Signing Variance}
\begin{figure}[t]
\centering
\includegraphics[width=\columnwidth]{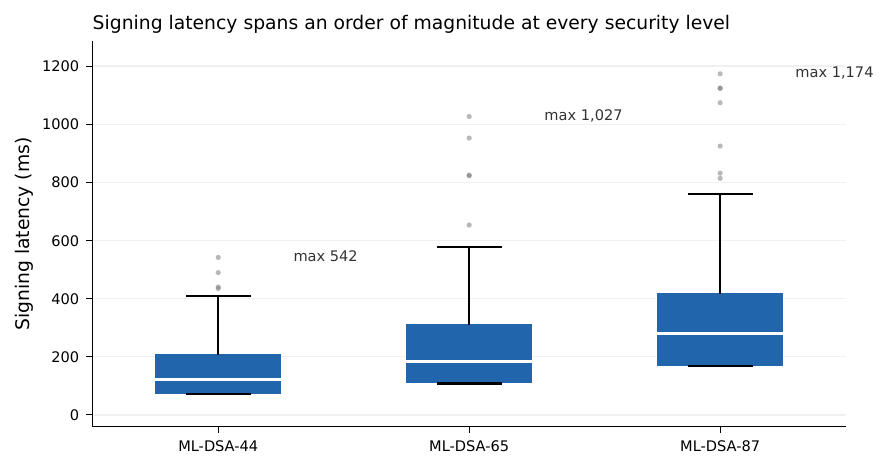}
\caption{\mldsa{} signing latency variance across security
  levels (campaign~A, $n{=}100$ per level).  Boxes span the
  interquartile range, whiskers the 1.5\,IQR fence; points
  beyond are high-iteration rejection-sampling events.  The
  annotated maxima are sample maxima at $n{=}100$ and understate
  the true tail---see Table~\ref{tab:tail}.}
\label{fig:variance}
\end{figure}

Table~\ref{tab:mldsa} presents \mldsa{} timing results.
Key generation and verification are deterministic (CV\,$<$\,1\%);
Figure~\ref{fig:variance} contrasts this with the spread of the
signing operation.
Signing, however, exhibits dramatic variance due to FIPS~204
rejection sampling.

ML-DSA-44 signing averages \SI{158.9}{\milli\second} but ranges
from \SI{70.1}{\milli\second} (best case: signature accepted on
first attempt) to \SI{541.7}{\milli\second} (worst case observed),
with a coefficient of variation of 67.8\%
and a 99th-percentile latency of \SI{489.4}{\milli\second}.
ML-DSA-65 averages \SI{256.6}{\milli\second} (CV\,=\,73.5\%,
p99\,=\,\SI{952.8}{\milli\second}), and ML-DSA-87 averages
\SI{355.2}{\milli\second} (CV\,=\,66.0\%,
p99\,=\,\SI{1124.5}{\milli\second}).

These quantiles are estimated from $n{=}100$ and are best read as
lower bounds.  A p99 from 100 samples is the single largest
order statistic below the maximum, and the campaign~B data show
directly how much that costs: re-running \mldsa{}-44 signing at
\num{10000} signatures raises p99 from \SI{489.4}{\milli\second} to
\SI{584.2}{\milli\second} (\SI{+19.4}{\percent}) and the observed
maximum from \SI{541.7}{\milli\second} to
\SI{1156.8}{\milli\second} ($2.1\times$).  We did not run
\num{10000}-signature campaigns at \mldsa{}-65 or 87, so their
tails are very likely heavier than the figures above rather than
lighter.  Because the rejection loop is geometric, the tail has no
finite worst case at any sample size; a deadline analysis should
use the model, not a measured maximum.  All deadline conclusions
in Section~\ref{sec:discussion} are stated in a direction that
this bias can only strengthen.

\begin{figure}[t]
\centering
\includegraphics[width=\columnwidth]{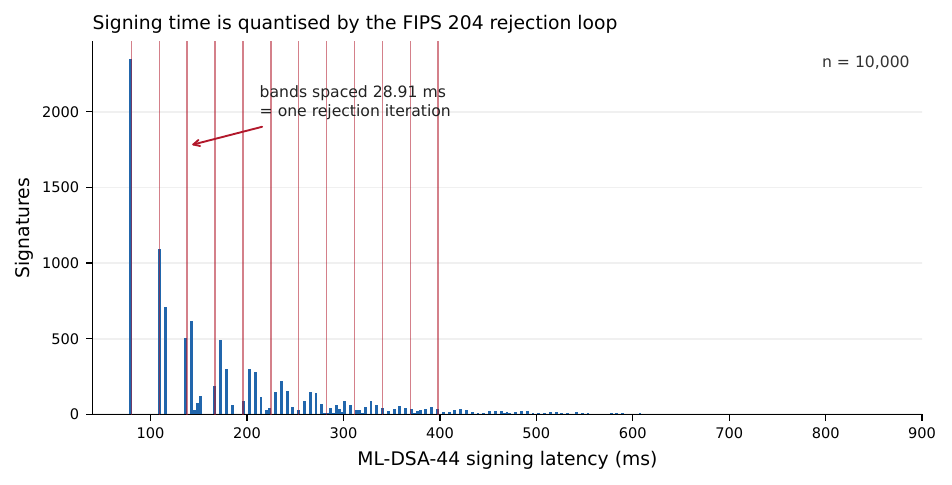}
\caption{\mldsa{}-44 signing time distribution over
  \num{10000} signatures (campaign~B).  The discrete bands are
  the FIPS~204 rejection loop: each band is one additional
  iteration, spaced by the fitted
  $c=\SI{28.91}{\milli\second}$ of
  Section~\ref{sec:latticefit}.  Band occupancy decays
  geometrically, and the tail extends to
  \SI{1156.8}{\milli\second}.}
\label{fig:histogram}
\end{figure}

This variance is an inherent property of the algorithm, not a
measurement artefact: the geometric-like distribution of signing
times (Fig.~\ref{fig:histogram}) matches the expected
rejection-sampling behaviour specified in FIPS~204.
The distribution also permits the per-iteration cost, and the
iteration count of each individual signature, to be recovered
without instrumenting the library.  Signing time is not
proportional to the iteration count: it is affine in it,
$t \approx t_0 + (k-1)c$, where $t_0$ is the one-off cost of
expanding the matrix~$\mathbf{A}$ and hashing the message and
$c$ is the marginal cost of one commit--challenge--reject
iteration.  Dividing the mean by the expected iteration count
therefore \emph{overestimates}~$c$ by absorbing~$t_0$.  We
develop this decomposition, and validate it against the
\num{10000}-signature sample, in
Section~\ref{sec:latticefit}; it explains why an iteration-count
bound alone does not bound signing time, a limitation recently
emphasised in the ML-DSA benchmarking-methodology
literature~\cite{pqshield2026pitfalls}.

\begin{table}[t]
\centering
\caption{ML-DSA timing on ARM Cortex-M0+ (RP2040, \SI{125}{\mega\hertz}).
  Signing: 100 runs; keygen/verify: 30 runs.
  Cycles = $t\,(\mu s) \times 125$.
  All quantiles in this paper are nearest-rank order statistics
  ($\lceil qn \rceil$-th smallest sample), never interpolated, so that
  every reported quantile is a latency that was actually observed;
  the same convention is used in Table~\ref{tab:tail}.}
\label{tab:mldsa}
\footnotesize
\setlength{\tabcolsep}{3pt}
\begin{tabular}{llrrrrr}
\toprule
\textbf{Variant} & \textbf{Op.} & \textbf{Mean} &
  \textbf{Cycles} & \textbf{CV} & \textbf{p95} & \textbf{p99} \\
 & & \textbf{(ms)} & \textbf{(kc)} & & \textbf{(ms)} & \textbf{(ms)} \\
\midrule
\multirow{4}{*}{\shortstack[l]{ML-DSA\\-44}}
  & KeyGen &  39.8 & \num{4979} &  0.9\% & --- & --- \\
  & Sign   & 158.9 & \num{19863} & 67.8\% & 383.4 & 489.4 \\
  & Sign\textsuperscript{\dag} & 182.7 & \num{22838} & 64.0\% & 421.3 & 584.2 \\
  & Verify &  44.0 & \num{5497} &  0.1\% & --- & --- \\
\midrule
\multirow{3}{*}{\shortstack[l]{ML-DSA\\-65}}
  & KeyGen &  68.4 & \num{8553} &  0.1\% & --- & --- \\
  & Sign   & 256.6 & \num{32077} & 73.5\% & 576.9 & 952.8 \\
  & Verify &  72.2 & \num{9026} &  0.0\% & --- & --- \\
\midrule
\multirow{3}{*}{\shortstack[l]{ML-DSA\\-87}}
  & KeyGen & 114.3 & \num{14287} &  0.5\% & --- & --- \\
  & Sign   & 355.2 & \num{44404} & 66.0\% & 831.9 & 1124.5 \\
  & Verify & 120.2 & \num{15028} &  0.0\% & --- & --- \\
\bottomrule
\multicolumn{7}{l}{\scriptsize \textsuperscript{\dag}Campaign~B,
  $n{=}\num{10000}$, \texttt{-Os} build: the best available estimate}\\
\multicolumn{7}{l}{\scriptsize of the \mldsa{}-44 signing tail.  Cite this row, not the
  $n{=}100$ row,}\\
\multicolumn{7}{l}{\scriptsize for tail quantiles; means are not comparable across
  optimisation}\\
\multicolumn{7}{l}{\scriptsize levels (Section~\ref{sec:replication}).  The $n{=}100$ rows
  understate $p_{99}$ by \SI{16}{\percent}.}\\
\end{tabular}
\end{table}

\subsection{Replication Campaign and Tail Behaviour}
\label{sec:replication}
The results above derive from a single measurement campaign
(campaign~A: 30~iterations for deterministic operations,
100~for signing).  To test whether those figures survive a change
of harness, toolchain, and optimisation level, and to resolve the
signing tail properly, we repeated the measurement as an
independent campaign~B on the same RP2040 unit with a separately
written harness, Pico~SDK~2.2.0, \texttt{-Os} pinned across every
translation unit, and sample sizes raised to \num{1000}
(deterministic) and \num{10000} (signing).
Campaign~B was originally captured for a separate multi-platform
study and was not conducted to support this paper, which removes
one source of experimenter bias.

\begin{table}[t]
\centering
\caption{Campaign~A (this paper, \texttt{-O3}, $n{=}30/100$)
  versus independent campaign~B (\texttt{-Os},
  $n{=}\num{1000}/\num{10000}$), same RP2040 unit.  $\Delta$ is the
  change in mean latency, with a 95\% interval propagated from the
  two campaigns' own CV and $n$ (normal approximation,
  $\mathrm{SE}=\mathrm{CV}/\sqrt{n}$, combined across campaigns).
  The interval bounds \emph{sampling} error only; the systematic
  difference between two builds is not a random quantity and no
  interval can be placed on it from two builds.  The two campaigns
  share no harness code and differ in SDK version, optimisation
  level, and sample size.}
\label{tab:replication}
\footnotesize
\setlength{\tabcolsep}{3pt}
\begin{tabular}{lrrrr}
\toprule
\textbf{Operation} & \textbf{A} & \textbf{B} &
$\boldsymbol{\Delta}$ \textbf{(95\% CI)} & \textbf{CV$_B$} \\
 & \textbf{(ms)} & \textbf{(ms)} & \textbf{(\%)} & \\
\midrule
\mlkem{}-512 KeyGen & 9.94   & 11.31  & $+13.7\pm0.60$ & 0.52\% \\
\mlkem{}-512 Encaps & 11.53  & 12.70  & $+10.1\pm0.07$ & 0.16\% \\
\mlkem{}-512 Decaps & 14.23  & 14.84  & $+4.3\pm0.07$  & 0.12\% \\
\mldsa{}-44 KeyGen  & 39.83  & 45.90  & $+15.2\pm0.38$ & 0.87\% \\
\mldsa{}-44 Sign    & 158.91 & 182.70 & $+15.0\pm15.3$ & 64.0\% \\
\mldsa{}-44 Verify  & 43.98  & 50.93  & $+15.8\pm0.04$ & 0.05\% \\
\bottomrule
\end{tabular}
\end{table}

\textbf{What replicates.}  Every qualitative conclusion in this
paper survives campaign~B unchanged: the ordering of operations,
the near-zero dispersion of all deterministic operations
(CV${\le}0.87\%$ in both campaigns), the confinement of variance
to signing alone, and the memory figures.  Peak stack agrees to
within \SI{32}{\byte} on \mlkem{}-512 (\num{9624} vs \num{9656})
and \SI{16}{\byte} on \mldsa{}-44 signing (\num{51824} vs
\num{51840}) despite the change of optimisation level, which is
expected: stack depth in these implementations is dominated by
statically sized polynomial arrays rather than by
compiler-allocated temporaries.

\textbf{What shifts.}  Absolute latencies rise by 4.3--15.8\%.
This is the optimisation level, not measurement noise: campaign~B
pins \texttt{-Os} across all translation units including the
shared \texttt{fips202} Keccak layer, whereas campaign~A inherits
\texttt{-O3}.  The spread across operations tracks how
SHAKE-dominated each one is---\mlkem{} decapsulation, which spends
proportionally the least time in Keccak, shifts least
($+4.3\%$).  The practical consequence is that absolute
millisecond figures in this paper are specific to a
speed-optimised build; a size-optimised firmware should budget
roughly 15\% more.  Relative comparisons, feasibility conclusions,
and the rejection-sampling analysis are unaffected.

\textbf{How much of that is sampling error?}  For the
deterministic operations, almost none; for signing, possibly all of
it.  The intervals in Table~\ref{tab:replication} answer only the
narrow question of whether two means could differ by chance.
Across the five deterministic operations the half-width is
\numrange{0.04}{0.60} percentage points against shifts of
\SIrange{4.3}{15.8}{\percent}: the shifts are resolved to two
significant figures and exclude zero by one to two orders of
magnitude.  \mldsa{}-44 signing is the exception.  Rejection
sampling gives campaign~A's $n{=}100$ sample a standard error of
\SI{6.8}{\percent} of its own mean, so the $+\SI{15.0}{\percent}$
shift carries an interval of $\pm\num{15.3}$ percentage points and
is, on its own, consistent with no change at all.  The
optimisation-level effect on signing is therefore inferred from the
deterministic operations of the same build---in particular
\mldsa{}-44 keygen and verify, which exercise the same
\texttt{fips202} and polynomial code and shift by
\SI{15.2}{\percent} and \SI{15.8}{\percent}---and not established
by the signing means.  For the ordering that
Section~\ref{sec:interunit} rests on, what matters is that even the
narrowest shift here, \mlkem{}-512 decapsulation at
$+\SI{4.3}{\percent}\pm\num{0.07}$, exceeds the largest inter-unit
difference (\SI{0.075}{\percent}) by a factor of \num{57}.

\begin{table}[t]
\centering
\caption{\mldsa{}-44 signing tail at $n{=}\num{10000}$
  (campaign~B) against the $n{=}100$ sample of campaign~A.
  Quantiles are reported only where the sample supports them:
  $n(1-q)\ge1$.}
\label{tab:tail}
\footnotesize
\begin{tabular}{lrr}
\toprule
\textbf{Statistic} & \textbf{A ($n{=}100$)} & \textbf{B ($n{=}\num{10000}$)} \\
\midrule
Mean      & 158.91 & 182.70 \\
Min       & 70.14  & 80.12 \\
Median    & 121.20 & 143.91 \\
p95       & 383.40 & 421.29 \\
p99       & 489.41 & 584.20 \\
p99.9     & ---    & 861.10 \\
Max       & 541.72 & 1156.79 \\
\bottomrule
\end{tabular}
\end{table}

\textbf{The tail is heavier than 100 samples suggested.}
Table~\ref{tab:tail} gives the point of the exercise.  The p99
estimated from 100 samples understates the \num{10000}-sample
value by 16\%, and the observed maximum more than doubles, from
\SI{541.7}{\milli\second} to \SI{1156.8}{\milli\second}---%
\num{14.4} times the fastest observed signature and
\num{8.0} times the median.  A 100-sample campaign cannot see
this: the worst case it observes is by construction close to its
p99, whereas the true distribution continues well past it.  Any
deployment sizing a watchdog or a real-time budget from
percentile statistics must therefore state the sample size the
percentile came from, and even \num{10000} samples bound only
what was observed, not what the standard admits.

\subsection{Inter-Unit Agreement}
\label{sec:interunit}
Campaigns~A and~B share a limitation: both ran on one physical
RP2040.  They bound sensitivity to harness, toolchain, and
optimisation level, but say nothing about manufacturing spread,
so every absolute figure in this paper rested on the assumption
that one unit represents the part.  Campaign~C tests that
assumption directly.

We measured a second, independently purchased RP2040 with the
identical firmware images used for campaign~C on the first unit,
and repeated the full sweep on both: \num{28} operations spanning
\mlkem{}, \mldsa{}, RSA-2048, ECDSA/ECDH-P256, and the on-chip
ring-oscillator RNG, at $n{=}\num{1000}$ for deterministic
operations, \num{2000} for RNG, and \num{100}/\num{3} for the two
slowest RSA operations.  The two captures total \num{58206} timed
runs.  Both dies report distinct hardware unique
IDs (\texttt{E663AC91D321B038} and \texttt{E66368254F75782F}) and
both report \SI{125}{\mega\hertz} \texttt{clk\_sys} read from the
hardware at boot, so the comparison is between physical parts and
not between clock configurations.

\textbf{Equivalence rather than difference testing.}  A
non-significant difference is not evidence of agreement, so we
pre-declared an equivalence margin of $\pm1\%$---below the
threshold at which any conclusion in this paper would change---and
tested each operation with two one-sided tests (TOST) at
$\alpha{=}0.05$, equivalent to requiring the 90\% confidence
interval on the relative difference to lie entirely inside the
margin.  Because a difference verdict from a small heavy-tailed
sample is unreliable, we additionally require $n{\ge}30$ per unit
and corroboration by a Mann--Whitney test before reporting any
operation as different; operations failing that power requirement
are reported as inconclusive rather than as agreement.  The three
\mldsa{} signing operations are rejection-sampled and dominated by
per-signature randomness, so they receive no verdict and appear in
the released data only.

\begin{figure}[t]
\centering
\includegraphics[width=\columnwidth]{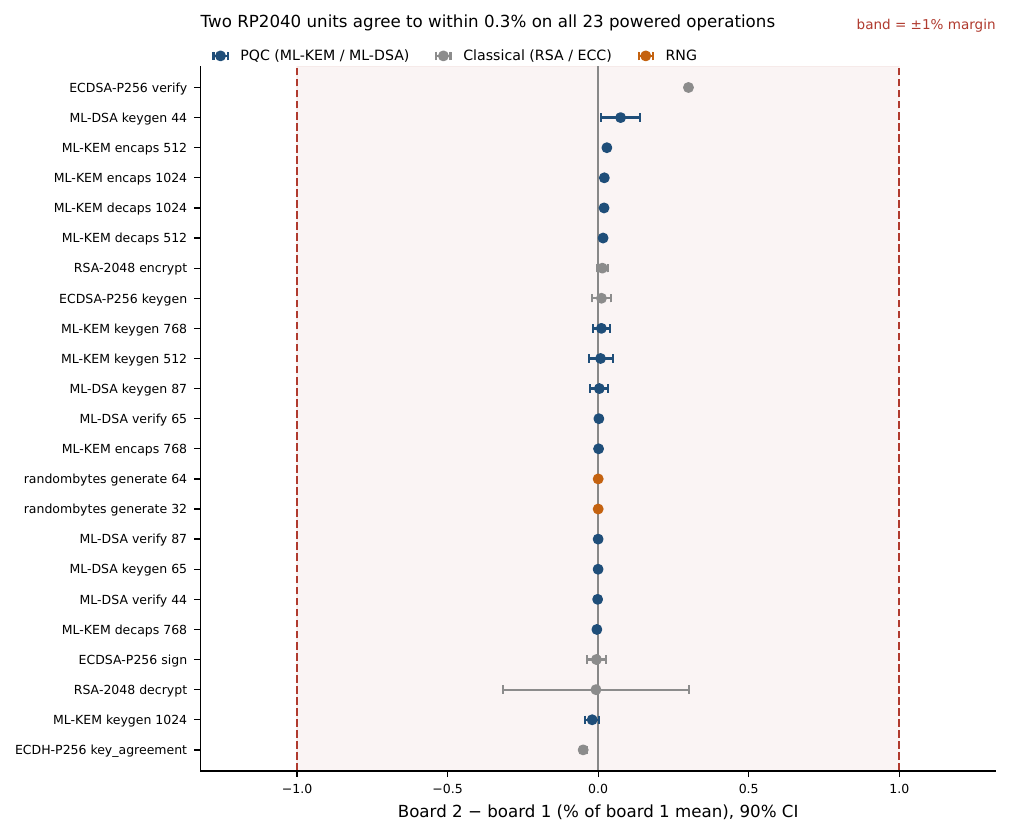}
\caption{Inter-unit agreement across the \num{23} adequately
  powered operations of campaign~C.  Points are the mean
  difference between two physical RP2040 units as a percentage of
  the first unit's mean; bars are 90\% confidence intervals, the
  interval TOST requires to fall inside the pre-declared $\pm1\%$
  equivalence margin (shaded).  Every operation clears the margin
  by at least a factor of three, and the \num{15} post-quantum
  operations clear it by more than an order of magnitude.}
\label{fig:interunit}
\end{figure}

\textbf{Result.}  All \num{23} adequately powered operations are
statistically equivalent within the margin
(Fig.~\ref{fig:interunit}); none is different.  The largest
observed inter-unit difference is \SI{0.300}{\percent}
(ECDSA-P256 verification) and every 90\% interval lies inside
$\pm\SI{0.315}{\percent}$.  Restricting to the \num{15}
post-quantum operations, which are what this paper is about, the
largest difference is \SI{0.075}{\percent}
(\mldsa{}-44 key generation) and the widest interval half-width is
\SI{0.140}{\percent}.  Nine operations agree to better than
\SI{0.005}{\percent}.  The RNG returns bit-identical timings on
both units at \num{32} and \num{64} bytes across \num{2000}
samples each.

\textbf{The family-wise error rate does not threaten this
conclusion.}  Twenty-three equivalence verdicts at
$\alpha{=}0.05$ inflate the chance that one is spurious, in the
unusual direction: false \emph{equivalence} rather than false
difference.  The largest TOST $p$-value across the \num{21}
operations for which a test statistic exists is \num{1.3e-7}
(RSA-2048 decryption), against a Bonferroni threshold of
$\alpha/23=\num{2.2e-3}$---four orders of magnitude of margin---and
the step-down Holm procedure rejects every null likewise.  Every
verdict here therefore holds at a family-wise $\alpha{=}0.05$; the
uncorrected per-test values are quoted only because correcting them
changes no verdict.  Appendix~\ref{sec:appendix-multiplicity} gives
the accounting, including why the two zero-dispersion RNG
operations are excluded from the family rather than counted as
passing.

\textbf{Two operations are reported as untested, not as
agreement.}  RSA-2048 key generation is rejection-sampled over
prime candidates and was measured at $n{=}3$ per unit, where the
\SI{90}{\percent} interval spans hundreds of percent; a
permutation test over all \num{20} partitions of $3{+}3$ samples
has a minimum attainable $p$ of \num{0.100}, so no difference can
reach significance at that sample size regardless of the data, and
both unit means fall inside the range campaign~A observed on a
single unit (\SIrange{46}{303}{\second} over five runs).
Resolving even a \SI{2}{\percent} effect at the dispersion that
campaign observed ($\mathrm{CV}\approx\SI{49}{\percent}$) would
require of order \num{9000} samples per unit at \SI{80}{\percent}
power, some \num{550} hours of board time per unit, which is not
proportionate for a baseline from which this paper draws no
conclusions.  Single-byte RNG generation is quantised to a
\SI{1}{\micro\second} timer at a mean of \SI{1}{\micro\second},
so its interval is set by timer resolution rather than by any
property of the hardware.  Both are reported as inconclusive.

\textbf{Interpretation.}  The comparison that matters is against
the effects this paper already documents, and it resolves into a
clean ordering of three magnitudes over the \num{15} deterministic
post-quantum operations:
\begin{itemize}
  \item \emph{Physical die:} ${\le}\SI{0.075}{\percent}$
    (this section, identical firmware image on two units).
  \item \emph{Compiler version at fixed \texttt{-O3}:}
    \SIrange{1.4}{2.5}{\percent}, mean \SI{1.9}{\percent}
    (campaign~A's \texttt{arm-none-eabi-gcc}~12.2.1 against
    campaign~C's~15.2.1).
  \item \emph{Optimisation level:} \SIrange{4.3}{15.8}{\percent}
    (campaign~B at \texttt{-Os},
    Section~\ref{sec:replication}).
\end{itemize}
A compiler-version bump therefore matters roughly \num{26} times
more than which unit ran the binary, and an optimisation-level
change between \num{57} and \num{211} times more.  The build
dominates the silicon.  That converts the build-sensitivity caveat
of Section~\ref{sec:replication} from a methodological assertion
into a measured ordering, and retires the single-unit limitation
for the quantities this paper reports.

We claim no more than this.  Two units purchased together are
likely to share a wafer lot, so this result bounds inter-unit
variation under realistic procurement rather than characterising
silicon variation across process corners, and it does not address
temperature or supply-voltage dependence, both held at laboratory
ambient.  What it establishes is that the single-unit figures
reported throughout this paper are not an artefact of the
particular die that produced them.

\subsection{Recovering Iteration Counts from Wall-Clock Time}
\label{sec:latticefit}
The \num{10000}-signature sample lets us do more than describe a
distribution: it lets us recover, for each individual signature,
how many rejection-sampling iterations produced it---using only a
\SI{1}{\mega\hertz} wall clock and no instrumentation of the
signing routine.

\textbf{Model.}  FIPS~204 signing repeats a candidate-generation
loop until a candidate passes the norm checks.  If each iteration
costs a fixed $c$ and the per-signature fixed work costs $t_0$,
then a signature accepted on iteration $k$ takes
$t \approx t_0 + (k-1)c$.  Rejections that fail the first norm
check exit early, contributing a smaller quantum $f$.  We
therefore model each observation as
\begin{equation}
t_i \;=\; t_{\min} + a_i c + b_i f + \varepsilon_i,
\qquad a_i \in \mathbb{Z}_{\ge 0},\; b_i \in \mathbb{Z}_{\ge 0},
\label{eq:lattice}
\end{equation}
anchoring $t_{\min}$ at the fastest observed signature, which by
construction is a single-iteration one.

\textbf{Fitting.}  For a candidate $(c,f)$ the integers are
determined greedily---$a_i=\lfloor (t_i-t_{\min})/c \rfloor$, then
$b_i$ by rounding the remainder to the nearest multiple of $f$,
with carry into $a_i$ when the remainder rounds up to a full
$c$---so the loss depends on $(c,f)$ alone.  We minimise the
root-mean-square residual
$\mathcal{L}(c,f) = \big(\tfrac{1}{n}\sum_i \varepsilon_i^2\big)^{1/2}$
by exhaustive grid search over
$c \in [\SI{28}{},\SI{30}{}]$\,ms and
$f \in [\SI{5.5}{},\SI{7.5}{}]$\,ms at \SI{10}{\micro\second}
resolution---a grid rather than a gradient method because
$\mathcal{L}$ is piecewise constant in the integers and therefore
not differentiable.
These bounds are themselves an assumption, and an important one:
Section~\ref{sec:latticegen} shows that widening the search
reveals the objective to be degenerate, so the window above is
doing part of the identification work.  We therefore report the
windowed fit here for continuity with the modal structure of
Figure~\ref{fig:histogram}, and in Section~\ref{sec:latticegen}
re-derive the same step from an uninformed search under a
criterion that does not presuppose the answer.

\textbf{Result.}  The optimum is
$c = \SI{28.91}{\milli\second}$,
$f = \SI{5.91}{\milli\second}$,
$t_{\min} = \SI{80.12}{\milli\second}$, with a residual of
\SI{0.27}{\milli\second}---\SI{0.149}{\percent} of the mean
signing time, and close to the
$\pm\SI{1}{\micro\second}\times$quantisation floor of the timer
once accumulated over a multi-iteration signature.  Resampling
the \num{10000} observations with replacement (200~replicates,
refitting each) gives
$c = 28.910 \pm \SI{0.001}{\milli\second}$ and
$f = 5.910 \pm \SI{0.003}{\milli\second}$: the parameters are
determined to the microsecond by the sample.

\textbf{Validation.}  The fit is not merely low-residual; the
integers it recovers carry the right statistics.  The mean
recovered iteration count is \num{4.34}, against the value
$\mathbb{E}[k]=4.25$ that FIPS~204 predicts for \mldsa{}-44---a
\SI{2.2}{\percent} deviation, from a quantity the fit was never
told.  The recovered distribution is geometric to the eye and to
a $\chi^2$ test over $k=1\ldots15$ plus a tail bin
($\chi^2=30.6$, 15~d.o.f.), and the single-iteration fraction is
\SI{23.5}{\percent} against the predicted $1/4.25 =
\SI{23.5}{\percent}$.  The $\chi^2$ above is computed against the FIPS-204 predicted
success probability $\hat{p}=1/4.25$ rather than against $\hat{p}$
estimated from the recovered counts.  Re-running the test with the
estimated parameter, which is the appropriate comparison for a
goodness-of-fit test on fitted data, gives $\chi^2=22.64$ on
14~d.o.f.\ ($p=0.066$; one degree of freedom is spent estimating
$\hat{p}$ from the sample): the geometric \emph{form} is not
rejected at the \SI{5}{\percent} level.  The original rejection
therefore reflects the small upward bias in $\mathbb{E}[k]$
reported above, not a failure of the geometric model.  We fix
$\alpha=0.05$ as our threshold throughout, so $p=0.066$ is a
non-rejection by that criterion.

A non-rejection is weak evidence on its own, so we ran three
further checks, reported in full in
Appendix~\ref{sec:appendix-latticestats}: a parametric bootstrap
that removes the asymptotic assumption ($p=0.069$), a
Kolmogorov--Smirnov check calibrated the same way ($p=0.50$), and
a power analysis against a beta-geometric alternative in which
acceptance probability drifts with attempt number.  The power
analysis is the one that bounds the claim: at $n=\num{10000}$ the
test reaches \SI{80}{\percent} power against departures that
inflate $\mathbb{E}[k]$ by \SI{4.4}{\percent}, so this campaign
rules out deviations that would move mean signing latency by more
than about \SI{5}{\percent} and cannot rule out smaller ones.  We
also confirmed by simulation that quantisation of the early-abort
step does not misassign signatures between adjacent $k$ bins at
the observed noise level.

We therefore claim only this: the geometric law is adequate at
the precision that matters for latency budgeting, not that it is
exact.  A deviation too small for us to detect is also too small
to change any timing figure in this paper.
The longest signature observed corresponds to $k=38$.

\subsection{Does the fit generalise across security levels?}
\label{sec:latticegen}
The decomposition above is fitted on \mldsa{}-44, where the
\num{10000}-signature campaign resolves $c$ and $f$.  Before using
the result we must confront a problem with the fit itself, which we
believe is the central methodological issue in this section and
which applies to any attempt to invert timing onto a lattice.

\begin{figure}[t]
\centering
\includegraphics[width=\columnwidth]{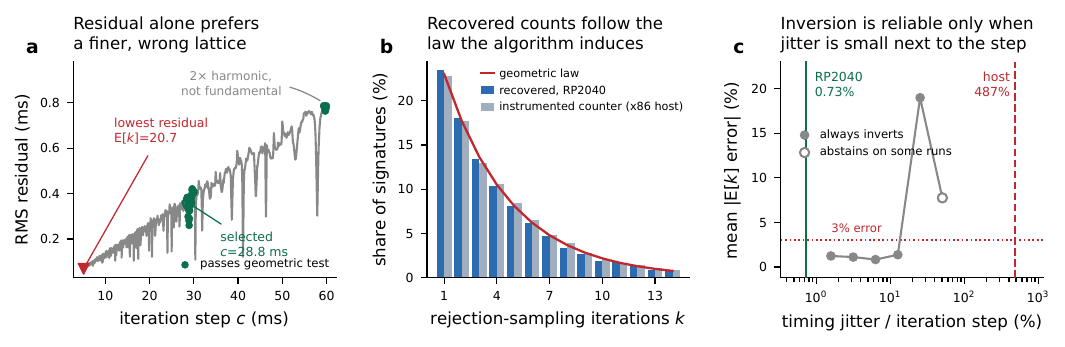}
\caption{Identifiability and validation of the iteration-recovery
  method.  (a)~The least-squares objective on the
  \num{10000}-signature \mldsa{}-44 sample is degenerate: residual
  declines toward finer lattices and the global minimum
  ($c=\SI{5.1}{\milli\second}$, $\mathbb{E}[k]=20.7$) is wrong.
  Points passing the geometric criterion are marked; the selected
  fundamental is $c=\SI{28.8}{\milli\second}$.  (b)~Iteration
  counts recovered on the RP2040 against the geometric law and
  against an instrumented rejection counter.  The two empirical
  series come from different platforms and are not a per-signature
  comparison: the recovered counts are RP2040 measurements, whereas
  the instrumented counts are \num{10000} signatures per level
  captured on an x86 host, the only build in which we could record
  ground-truth $k$.  They are shown together because both should
  follow the same algorithm-induced law.  (c)~Reliability regime: mean
  absolute error in $\mathbb{E}[k]$ against timing jitter expressed
  as a fraction of one iteration step.  Open markers denote
  configurations where the method abstained on some runs.  The
  RP2040 and the x86 host fall on opposite sides of the usable
  region.}
\label{fig:recovery}
\end{figure}

\textbf{The least-squares objective is degenerate.}  Fitting
$t \approx t_0 + (k-1)c + \varepsilon$ by minimising RMS residual
over $(c,f)$ does not identify $c$, and the failure is structural
rather than statistical.  Any $c' = c/m$ for integer $m$ also
places every observation near a lattice point, and finer lattices
fit strictly better: as $c \to 0$ the residual goes to zero for
\emph{any} data whatsoever.  Scanning $c$ over a wide uninformed
grid on the \num{10000}-signature \mldsa{}-44 sample, the global
residual minimum lies at $c=\SI{5.1}{\milli\second}$ with
$f=\SI{0.255}{\milli\second}$, giving RMS
\SI{0.069}{\milli\second}---a \emph{better} residual than the
value we report---and implying $\mathbb{E}[k]=20.7$, roughly five
times the FIPS-204 expectation.  Reporting the residual minimiser
would therefore have produced a confidently wrong answer.
Figure~\ref{fig:recovery}(a) shows the objective: the residual
curve declines monotonically toward small $c$, with harmonic
aliases of the true step visible as local minima.

\textbf{Identification requires a criterion outside the residual.}
We take it from the algorithm.  FIPS~204 signing repeats an
independent Bernoulli trial until acceptance, so the recovered
iteration counts must follow a geometric distribution; a spurious
lattice at $c/m$ splits each true iteration class into $m$
sub-classes and destroys that shape.  We therefore admit only
those $c$ whose recovered counts pass a $\chi^2$ goodness-of-fit
test against a geometric law ($p>0.05$, parameter estimated from
the recovered counts), and among admissible candidates select the
\emph{fundamental}---the smallest $c$ that is not itself an
integer multiple of a smaller admissible value.  On an uninformed
grid this returns $c=\SI{28.8}{\milli\second}$,
$f=\SI{1.44}{\milli\second}$, $\mathbb{E}[k]=4.34$
($\chi^2$ $p=\num{0.115}$), with no admissible candidate below
\SI{28}{\milli\second}.  A bootstrap over signatures gives a
\SI{90}{\percent} interval for $\mathbb{E}[k]$ of
$[\num{4.28},\num{4.37}]$ and places $c$ in
\SIrange{28.5}{29.8}{\milli\second} in every resample.  The
recovered $\mathbb{E}[k]=4.34$ sits \SI{2.1}{\percent} above the
FIPS-204 expectation of 4.25.
This is the substantive check on Section~\ref{sec:latticefit}:
the uninformed search recovers the same iteration step
($\SI{28.8}{\milli\second}$ on a coarser grid, against
$\SI{28.91}{\milli\second}$ from the windowed fit) and the same
mean iteration count, without being given the window.  The
windowed fit's finer estimate of $f$ is not recovered here, as the
uninformed grid is too coarse to resolve it; $f$ is a
second-order term and none of our conclusions depend on it.

\textbf{Validation against an instrumented counter.}  A criterion
justified by theory still needs checking against truth, so we
patched the PQClean signing routine with a rejection-loop counter
and captured \num{10000} instrumented signatures at each of the
three security levels on an x86 host, recording the true iteration
count alongside each timing.  The counter confirms the model the
method assumes: median signing time is linear in the true
iteration count, with $R^2$ of \numrange{0.994}{0.999} across the
three levels, fitting per-$k$ medians over those iteration counts
observed at least 20 times.  We state that inclusion rule because
the result is sensitive to it: admitting sparse tail bins down to
five observations lowers the range to \numrange{0.958}{0.995},
a median over five samples being itself noisy.  Per-\emph{sample}
$R^2$ is far lower (\numrange{0.09}{0.77}), because host scheduling
jitter on x86 is large relative to one iteration---the same effect
that makes the method abstain there, quantified below.  The true
iteration counts are geometric with $\mathbb{E}[k]$ of \num{4.386},
\num{5.181} and \num{3.939}---within \SIrange{1.6}{3.2}{\percent}
of the FIPS-204 values.  Figure~\ref{fig:recovery}(b) compares the distribution
recovered on the RP2040 against both the geometric law and the
instrumented counts.  The instrumented data and the patch are
released with the artefact.

\textbf{When the method works, and when it must abstain.}  The
governing quantity is timing jitter measured against the iteration
step.  On the RP2040 the per-iteration step is
\SI{28.8}{\milli\second} and the residual jitter is
\SI{0.73}{\percent} of it; on the x86 host the same algorithm runs
so much faster that jitter is \SI{487}{\percent} of a
\SI{0.082}{\milli\second} step, and the lattice is unresolvable in
principle.  A controlled sweep (Figure~\ref{fig:recovery}(c))
locates the transition: mean absolute error in $\mathbb{E}[k]$
stays below \SI{2}{\percent} while jitter is under about
\SI{13}{\percent} of the step, then degrades sharply.  Applied
blind to the instrumented host data, the geometric criterion
\emph{abstains} at levels 44 and 65---no candidate lattice passes
the test---rather than returning a wrong answer, and at level 87,
where jitter is relatively smaller, it returns $\mathbb{E}[k]=4.85$
against a true \num{3.94}, an error of \SI{23}{\percent}.  This
abstention behaviour is the practically important property: the
method reports that it cannot answer rather than answering
incorrectly.

\textbf{Scope of the claim.}  The recovery result is established at
\mldsa{}-44 on the \num{10000}-signature campaign.  We do not claim
it at \mldsa{}-65 or 87 on hardware, because we did not run
\num{10000}-signature campaigns at those levels and the $n=100$
samples available there are far too small: the same procedure on
$n=100$ returns bootstrap intervals for $\mathbb{E}[k]$ wide enough
to contain most of the plausible range.  The one cross-level
observation we make is that the coarse modal spacing grows with the
parameter set (roughly 29, 43 and \SI{50}{\milli\second} at levels
44, 65 and 87), consistent with a per-iteration cost scaling with
the matrix dimension; this speaks to $c$, not to the iteration
count.  Appendix~\ref{sec:appendix-latticestats} reports the
goodness-of-fit diagnostics, and
\texttt{scripts/verify\_identifiability.py} in the artefact
regenerates every number in this subsection, including the
degenerate minimiser we rejected.

\textbf{Why it matters.}  Separating $t_0$ from $c$ answers a
question the aggregate mean cannot.  Of the
\SI{80.12}{\milli\second} minimum signature,
\SI{51.21}{\milli\second}---\SI{64}{\percent}---is fixed setup
that a caller pays once regardless of how the rejection loop
turns out, and only \SI{28.91}{\milli\second} is the marginal
cost of an additional attempt.  Dividing the \num{10000}-signature
mean of \SI{182.70}{\milli\second} by the FIPS-204
$\mathbb{E}[k]=4.25$, the obvious estimate, gives
\SI{43.0}{\milli\second} per iteration and overstates the true
marginal cost by \SI{49}{\percent}.  The distinction is what
determines whether an implementer should attack the setup path
or the loop body.  The method needs no cycle counter, no
debugger, and no modification to the implementation under test.
It is therefore applicable in principle to any rejection-sampling
scheme on a platform with a microsecond-resolution clock, subject
to the two conditions established above: the sample must be large
enough (\num{10000} signatures sufficed at \mldsa{}-44 and 100 did
not), and timing jitter must be small relative to one iteration.

\subsection{Security Implications of Iteration-Count Recovery}
\label{sec:sidechannel}
What we have just described is a timing side channel, and it
should be labelled as one rather than presented purely as a
measurement convenience.  The method takes an uninstrumented
wall clock and recovers a secret-dependent quantity---the number
of rejection-sampling iterations that produced a given
signature---without touching the implementation.  That places it
in the same family as the attacks on Fiat--Shamir-with-aborts
signatures, and the relationship deserves stating precisely,
and because it qualifies the scope of the side-channel
limitation stated in Section~\ref{sec:limitations}.

\textbf{What the recovered quantity is.}  The iteration count
$k$ is a function of the secret key only through the rejection
condition: signing repeats until a candidate passes the norm
checks, and how many attempts that takes depends on the norms of
intermediate values that involve $\mathbf{s}_1,\mathbf{s}_2$.  It
is not itself a nonce or a key.  Under FIPS~204's parameters $k$
is close to geometric with the scheme's designed acceptance
rate---the fit in this section confirms as much at
\mldsa{}-44---so a single observation carries at most a few bits,
and those bits are drawn from a distribution the specification
publishes in advance.

\textbf{Relation to prior attacks.}  The rejection-sampling step
is known leakage territory.  Bootle et
al.~\cite{bootle2018ilwe} showed that the linear leakage from
that step in BLISS, initially thought unexploitable, yields key
recovery once it is cast as learning-with-errors without modular
reduction.  For ML-DSA specifically, Zhou et
al.~\cite{zhou2025rejected} recover the private key from leakage
of \emph{rejected} signatures' challenges by reduction to integer
linear programming.  Migliore et al.~\cite{migliore2019masking} treat the
rejection step as sensitive when masking Dilithium.

The distinction that matters here is between the \emph{count} of
iterations and the \emph{content} of the rejected attempts.  The
published key-recovery results exploit content: the challenge
polynomial of a rejected attempt, its Hamming weight, or a
quadratic function of the secret evaluated during rejection.  Our
method observes only how many attempts occurred, and observes it
at a granularity of one iteration---roughly
\SI{29}{\milli\second} at \mldsa{}-44---with no visibility into
any intermediate value.  We are therefore not claiming a key
recovery, and we did not attempt one.

\textbf{What we can and cannot conclude.}  We make the negative
claim only in the form we can support: an adversary who learns
$k$ for each of many signatures learns a sample from a
distribution whose parameter is fixed by the standard, and we are
not aware of a published reduction from iteration counts alone to
key or nonce recovery.  We emphasise that this is an absence of a
known attack, not a proof of its impossibility, and the two
should not be confused.  Two considerations argue for caution
rather than reassurance.  First, the BLISS history is precisely a
case where leakage was dismissed as unexploitable and was later
exploited once the right algebraic framing was
found~\cite{bootle2018ilwe}.  Second, $k$ is not independent of
the secret across signatures; a deployment that signs many
messages under one key emits a correlated stream of these
counts, and the security of that stream is a question for
cryptanalysis rather than for a benchmarking paper.

\textbf{Practical consequence.}  A device whose signing latency
is externally observable---over a network, a bus, or a power
rail---is already exposing $k$ to anyone who can time it, at far
less effort than a power or EM campaign requires, and our results
quantify how cleanly: at \mldsa{}-44 the iteration quantum is
\SI{28.91}{\milli\second} against a residual standard deviation
of \SI{0.27}{\milli\second}, a separation of roughly
$100\times$.  This does not depend on our method being used; it
is a property of the implementation's timing profile.
Implementations for which iteration-count leakage is in scope
should therefore constant-time the signing loop or add
latency randomisation, and the cost of doing so belongs in the
overhead budget alongside masking.  We flag this as a
consequence of our measurements for implementers, and as an open
question for cryptanalysts, rather than as a result of our own.

\textbf{Disclosure.}  We made no coordinated disclosure to
PQClean, NIST, or implementers before submitting this paper, and
we state that plainly rather than leave it to inference.  Our
reasoning is that the finding does not meet the threshold that
embargo exists to serve.  The quantity recovered is the number of
rejection-sampling iterations, whose distribution FIPS~204
specifies in advance; no key, nonce, or intermediate value is
recovered, and no key-recovery attack is claimed or attempted.
The data-dependent latency of rejection sampling is a documented
property of the Fiat--Shamir-with-aborts design---it is the
reason the specification discusses constant-time
considerations---rather than a defect introduced by the PQClean
implementation, so there is no implementation bug for a
maintainer to fix and no fix for a user to wait for.  Readers who
consider any secret-dependent timing channel reportable are
entitled to weigh that differently, and the measurements needed
to do so are in Section~\ref{sec:latticefit} and our released
data.  Should a reduction from iteration counts to key material
be demonstrated, the appropriate response would be disclosure to
implementers at that point, and we would treat it as such.

\subsection{Classical Baselines}
Table~\ref{tab:classical} presents classical algorithm timings
on the same RP2040 hardware.
The baselines are chosen for \emph{classical} security-level
alignment, and the alignment is deliberately conservative rather
than exact.  RSA-2048 and P-256 both provide roughly 112--128
bits of classical security, placing them against ML-KEM-512 and
ML-DSA-44 (NIST category~1, comparable to AES-128).  This is the
weakest post-quantum parameter set, so the comparison is the
least favourable one available to ML-KEM and ML-DSA: the
higher-security post-quantum variants in
Tables~\ref{tab:mlkem}--\ref{tab:mldsa} have no deployed
classical counterpart on this hardware, since RSA-3072 and
P-384 would be slower still.
Against a \emph{quantum} adversary the classical baselines
provide no security at all, so the comparison should be read as
the cost of migration from today's deployed primitives, not as a
like-for-like security trade.
RSA-2048 key generation is exceptionally slow (mean
\SI{215.6}{\second}, individual runs ranging from 46.1 to
\SI{303.0}{\second}) due to the probabilistic prime-finding
process on a processor without hardware acceleration.
Only $n=5$ runs were collected (total wall-clock time
${\sim}$18\,min), yielding a wider confidence interval than
the 30-run PQC benchmarks; the high CV of 48.6\% (sample
standard deviation, $n{-}1$ denominator, as throughout) reflects
the inherent variability of prime search rather than
measurement noise.
RSA-2048 decryption requires \SI{3393}{\milli\second}---two
orders of magnitude slower than ML-KEM decapsulation.
A complete ECDH\,P-256 key agreement---comprising both parties'
key generation, public-value exchange, and shared-secret
computation---takes \SI{617.0}{\milli\second} (\num{77128}\,kc).
By contrast, a full ML-KEM-512 handshake (keygen + encaps + decaps)
completes in \SI{35.7}{\milli\second} (\num{4464}\,kc),
approximately $17\times$ faster for establishing a shared secret.
We note that this comparison is between an IND-CCA2 KEM and a
raw Diffie--Hellman key agreement; in practice, ECDH is typically
wrapped in an authenticated key exchange (e.g., ECDH-ECDSA in TLS),
which would add further overhead to the classical side.

\begin{table}[t]
\centering
\caption{Classical cryptography timing on ARM Cortex-M0+
  (RP2040, \SI{125}{\mega\hertz}).  30 runs per operation
  (RSA-2048 keygen: 5 runs due to ${\sim}$216\,s mean).
  \emph{No comparative claim in this paper rests on the
  \colorbox{black!8}{shaded} RSA-2048 keygen row}: it is
  shaded because it is the sole $n{=}5$ row in a table of
  $n{=}30$ rows, and is reported for order-of-magnitude
  context only.  The migration comparisons in the abstract and
  discussion use the $n{=}30$ ECDH and ECDSA rows.
  Cycles derived as $t\,(\mu s) \times 125$.}
\label{tab:classical}
\footnotesize
\setlength{\tabcolsep}{3pt}
\begin{tabular}{llrrc}
\toprule
\textbf{Algorithm} & \textbf{Op.} & \textbf{Mean} &
  \textbf{Cycles} & \textbf{CV} \\
 & & \textbf{(ms)} & \textbf{(kc)} & \\
\midrule
\multirow{3}{*}{RSA-2048}
  & \cellcolor{black!8}KeyGen\textsuperscript{\dag}
    & \cellcolor{black!8}\num{215612}
    & \cellcolor{black!8}\num{26951438}
    & \cellcolor{black!8}48.6\% \\
  & Encrypt & 139.7 & \num{17460} & 1.2\% \\
  & Decrypt & \num{3393} & \num{424112} & 2.4\% \\
\midrule
\multirow{3}{*}{\shortstack[l]{ECDSA\\P-256}}
  & KeyGen & 83.6  & \num{10455} & 0.4\% \\
  & Sign   & 92.6  & \num{11575} & 0.4\% \\
  & Verify & 321.4 & \num{40170} & 0.0\% \\
\midrule
ECDH P-256\textsuperscript{*}
  & Key Agr. & 617.0 & \num{77128} & 0.4\% \\
\bottomrule
\multicolumn{5}{l}{\scriptsize \textsuperscript{\dag}$n{=}5$, CV\,$=$\,48.6\%: indicative
  only, not comparable in}\\
\multicolumn{5}{l}{\scriptsize \phantom{\textsuperscript{\dag}}reliability to the
  $n{=}30$ rows.}\\
\multicolumn{5}{l}{\scriptsize Cycles are kilocycles throughout,
  $t\,(\mu s)\times125$; RSA-2048 keygen is}\\
\multicolumn{5}{l}{\scriptsize \num{26951438}\,kc, i.e.\ ${\sim}$27 gigacycles.}\\
\multicolumn{5}{l}{\scriptsize \textsuperscript{*}Full
  handshake: \texttt{ecdh\_make\_params} (A keygen),}\\
\multicolumn{5}{l}{\scriptsize \phantom{\textsuperscript{*}}%
  \texttt{ecdh\_make\_public} (B keygen),
  2$\times$\texttt{ecdh\_calc\_secret}.}
\end{tabular}
\end{table}

\subsection{Energy Consumption}
Because the datasheet model assumes constant active power, energy
here is a unit conversion of the timing results
(\SI{79.2}{\micro\joule\per\milli\second}) and carries no
information beyond Tables~\ref{tab:mlkem} and~\ref{tab:mldsa}.  We
state the headline figures only, because millijoules connect
directly to battery capacity in a way milliseconds do not: a
complete \mlkem{}-512 key exchange costs an estimated
\SI{2.83}{\milli\joule} against \SI{48.9}{\milli\joule} for a full
ECDH\,P-256 key agreement ($17\times$), and \mldsa{}-44
keygen~+~sign~+~verify costs \SI{19.2}{\milli\joule} against
\SI{39.4}{\milli\joule} for the equivalent ECDSA\,P-256 cycle.  The
per-operation table is given in
Appendix~\ref{sec:appendix-energy}.  These are estimates from a
datasheet power model, not measurements; the ratios hold under any
constant-power assumption, the absolute values do not.

\subsection{Memory Footprint}
Table~\ref{tab:memory} presents peak stack usage (measured via
stack painting) and code size (from \texttt{arm-none-eabi-size})
for each algorithm variant.  Three column definitions apply.
\emph{Peak stack} is the stack-painting high-water mark for the
worst-case operation, measured with the harness itself excluded
(painting begins at the current SP inside the benchmark frame).
\emph{Code} is the \texttt{.text} contribution of the algorithm's
translation units from \texttt{arm-none-eabi-size -A}.
\emph{Total RAM} is peak stack \emph{plus} the caller-allocated
key-material buffers that the harness holds live across the
worst-case operation---for \mldsa{} the public key, secret key and
signature ($pk+sk+\sigma$), for \mlkem{} the public key, secret
key, ciphertext and both shared-secret buffers.  These are declared
in the benchmark frame and are therefore \emph{not} counted in the
Peak~stack column (Section~\ref{sec:methodology}).  Total~RAM
\emph{excludes} the interrupt stack, the USB-CDC/stdio buffers used
only to emit results, the Pico~SDK runtime, and
\texttt{.data}/\texttt{.bss} of the C library, none of which a
deployed signer or decapsulator needs.
Every KB figure is a campaign~A byte count divided by \num{1024}
and quoted to one decimal place (ML-KEM-512 decapsulation, for
instance, is \num{9624}\,bytes $=$ \SI{9.4}{\kilo\byte}); the byte
counts are in the released data, and campaign~B agrees to within
\SI{32}{\byte} despite the change of optimisation level
(Section~\ref{sec:replication}).

ML-KEM-512's peak stack usage during decapsulation is 9.4\,KB,
fitting within a 10\,KB allocation.  ML-KEM-768 and -1024
require 14.2\,KB and 20.0\,KB respectively for decapsulation.
ML-DSA is substantially more memory-intensive: ML-DSA-44 signing
peaks at 50.6\,KB, ML-DSA-65 at 77.6\,KB, and ML-DSA-87 at
119.6\,KB.  The steep increase reflects the larger
$K \times L$ polynomial vectors allocated on the stack by the
PQClean reference~C signing function.
Our benchmarks use a custom linker script allocating 192\,KB of
stack to accommodate these requirements.

Code size scales modestly with security level for ML-KEM
(5.1--6.7\,KB), while ML-DSA requires 8.4--8.9\,KB.
Both families share a common PQClean utility library
(\textasciitilde{}5\,KB of application-relevant code for hashing
and polynomial operations).

\begin{table}[t]
\centering
\caption{Memory footprint on RP2040 (264\,KB SRAM, 2\,MB flash).
  Column definitions, exclusions, and the rounding convention are
  given in the text above.  \textbf{Fits 264\,KB} indicates that
  the Total~RAM figure is below the RP2040's physical SRAM, i.e.\
  the variant is executable on the part at all; it is not a claim
  of headroom for an application or network stack.
  KB $=$ 1024\,bytes throughout.}
\label{tab:memory}
\footnotesize
\setlength{\tabcolsep}{3pt}
\resizebox{\columnwidth}{!}{%
\begin{tabular}{l l r r c r}
\toprule
 & & \multicolumn{3}{c}{\textbf{SRAM (KB)}} & \textbf{Flash} \\
\cmidrule(lr){3-5}
\textbf{Variant} & \textbf{Peak Op} & \textbf{Stack} & \textbf{Total} & \textbf{Fits} & \textbf{Code} \\
 & & & & \textbf{264\,KB} & \textbf{(KB)} \\
\midrule
ML-KEM-512  & decaps &  9.4 & 12.6 & \cmark & 5.1 \\
ML-KEM-768  & decaps & 14.2 & 18.9 & \cmark & 5.2 \\
ML-KEM-1024 & decaps & 20.0 & 26.2 & \cmark & 6.7 \\
\midrule
ML-DSA-44   & sign   & 50.6 & 56.8 & \cmark & 8.9 \\
ML-DSA-65   & sign   & 77.6 & 86.7 & \cmark & 8.4 \\
ML-DSA-87   & sign   & 119.6 & 131.5 & \cmark & 8.7 \\
\bottomrule
\end{tabular}
}
\end{table}

\subsection{Cortex-M0+ vs Cortex-M4 Comparison}
\begin{figure}[t]
\centering
\includegraphics[width=\columnwidth]{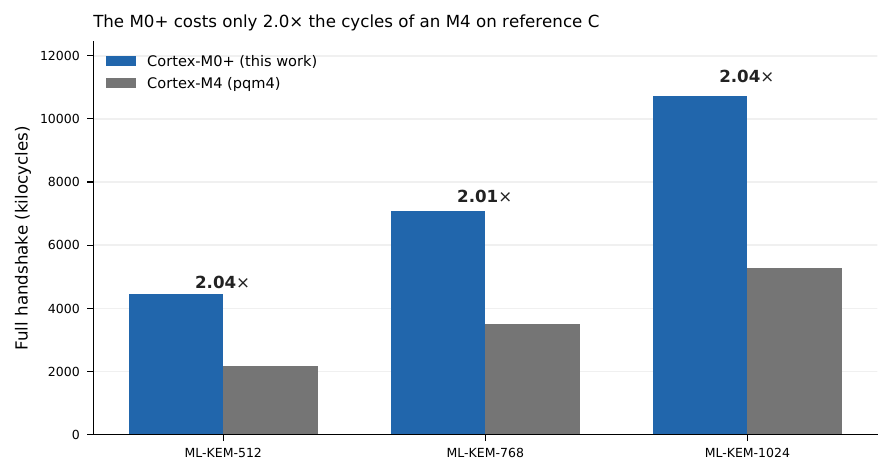}
\caption{\mlkem{} handshake cycle counts: \armmo{} (this work,
  \SI{125}{\mega\hertz}) vs Cortex-M4 (pqm4).  The \armmo{}
  incurs a 2.01--2.04$\times$ slowdown despite lacking
  \texttt{UMULL}, DSP, and SIMD instructions.  Both platforms
  are compared in cycles, so the ratio is independent of clock
  frequency.}
\label{fig:m0m4}
\end{figure}

Table~\ref{tab:m0m4} compares \mlkem{} handshake times on our
\armmo{} results with published pqm4 \armfour{} \emph{reference~C}
cycle counts from Kannwischer et al.~\cite{kannwischer2019pqm4} (pqm4 \texttt{clean} reference
implementations, not the optimised assembly variants; \texttt{benchmarks.csv}
at commit \texttt{90bfb630}, built with \texttt{arm-none-eabi-gcc} at
\texttt{-O3} and measured by that project on a NUCLEO-L4R5ZI at
\SI{24}{\mega\hertz}).
Two clock rates are in play, so we state the convention
before the numbers.  The M4 millisecond column divides the
published M4 \emph{cycle counts} by \SI{125}{\mega\hertz} (our
M0+ clock), \emph{not} by the \SI{24}{\mega\hertz} at which pqm4
measures them.  That column therefore answers ``how long would
the M4's work take if it ran at our clock'', which isolates the
architectural difference from the accident that the two boards
are clocked differently; it is not a claim about how fast a
NUCLEO-L4R5ZI runs.  The slowdown column needs no such
convention: it is a ratio of cycle counts and is independent of
both clock rates.
This normalisation shows how the M4 would perform at the same
clock speed, isolating the microarchitectural difference; the
same comparison is plotted in Figure~\ref{fig:m0m4}.  We note that the pqm4 optimised assembly
implementations are 2--4$\times$ faster than reference~C on M4;
the slowdown relative to \emph{optimised} M4 code would therefore
be proportionally larger (estimated 4--8$\times$).

The \armmo{} incurs a 2.01--2.04$\times$ slowdown across
security levels relative to M4 reference~C.
This is notably \emph{lower} than the 3--7$\times$ penalty
sometimes assumed for M0+ versus M4, likely because:
(i)~the RP2040's single-cycle multiplier mitigates the
absence of \texttt{UMULL};
(ii)~PQClean reference~C does not exploit M4-specific
instructions, so the M4 advantage is reduced when both platforms
run the same C code; and
(iii)~ML-KEM's NTT kernel is multiply-intensive, and the
dominant cost is 32-bit multiplication, which both platforms
execute in one cycle.

\begin{table}[t]
\centering
\caption{\mlkem{} full handshake: Cortex-M0+ vs Cortex-M4. M4 reference~C cycle counts are the \texttt{clean} implementations from the pqm4 benchmark set~\cite{kannwischer2019pqm4} (\texttt{benchmarks.csv} at commit \texttt{90bfb630}, 2025-05-22), measured by that project on a NUCLEO-L4R5ZI at \SI{24}{\mega\hertz} to avoid flash wait states. Both sides are reference~C at \texttt{-O3} (pqm4's default), so the comparison is flag- and implementation-matched; it is a cycle-count comparison and is independent of the differing clock rates. M4~(ms) converts M4 cycles at \SI{125}{\mega\hertz} to show what the same work would cost at the M0+'s clock.
Slowdown is the ratio of the two cycle-count columns, rounded
half-up to two decimals; the unrounded values are \num{2.043},
\num{2.013} and \num{2.035}, so the \numrange{2.01}{2.04} range
quoted in the text is the rounded envelope of ratios spanning
\numrange{2.013}{2.043}.}
\label{tab:m0m4}
\footnotesize
\setlength{\tabcolsep}{4pt}
\begin{tabular}{lrrrrr}
\toprule
\textbf{Variant} & \textbf{M0+} & \textbf{M0+} & \textbf{M4} & \textbf{M4} & \textbf{Slow-} \\
                 & \textbf{(ms)} & \textbf{(kc)} & \textbf{(ms)} & \textbf{(kc)} & \textbf{down} \\
\midrule
\mlkem{}-512  & 35.7 & \num{4464}  & 17.5 & \num{2185} & 2.04$\times$ \\
\mlkem{}-768  & 56.6 & \num{7077}  & 28.1 & \num{3515} & 2.01$\times$ \\
\mlkem{}-1024 & 85.7 & \num{10716} & 42.1 & \num{5265} & 2.04$\times$ \\
\bottomrule
\end{tabular}
\end{table}

\section{Discussion}
\label{sec:discussion}

\subsection{Does ML-KEM Fit on Class-1 IoT Devices?}
The RP2040's 264\,KB SRAM comfortably accommodates all \mlkem{}
variants.  Even the largest key object (ML-KEM-1024's
\num{3168}-byte secret key) represents only 1.2\% of available
SRAM.  However, stack requirements are non-trivial:
ML-KEM-512 decapsulation peaks at \num{9624} bytes in campaign~A
(\SI{9.4}{\kilo\byte}, fitting a \SI{10}{\kilo\byte} allocation;
campaign~B measured \num{9656} at \texttt{-Os},
Section~\ref{sec:replication}), while ML-KEM-1024 requires
\num{20488} bytes.
ML-DSA signing is far more demanding: ML-DSA-87's signing function
requires 119.6\,KB of stack (Table~\ref{tab:memory}), consuming
roughly 45\% of the RP2040's total SRAM for the call stack alone.
Deployments must ensure adequate stack allocation, particularly
for ML-DSA at higher security levels.
With a full handshake completing in \SI{35.7}{\milli\second}
(ML-KEM-512) to \SI{85.7}{\milli\second} (ML-KEM-1024),
ML-KEM is computationally feasible for Class-1 IoT key exchange
at the algorithm level; protocol overhead (TLS/DTLS handshake,
certificate chain) would add to the total latency in practice.

\subsection{Is ML-DSA Signing Variance Acceptable?}
The high coefficient of variation (66.0--73.5\%) in ML-DSA signing
is a significant concern for IoT applications with hard real-time
constraints.  The 99th-percentile signing time for ML-DSA-44 is
\SI{584.2}{\milli\second} at $n{=}\num{10000}$
(Table~\ref{tab:tail})---$7.3\times$ the fastest signature observed
in that campaign---with a maximum of \SI{1156.8}{\milli\second}.
For ML-DSA-87 the $n{=}100$ 99th-percentile already reaches
\SI{1124.5}{\milli\second} (over 1\,second); we ran no
\num{10000}-signature campaign at that level, so the true quantile
is higher still (Section~\ref{sec:results}).

For applications with 100\,ms deadlines (e.g., industrial
control loops), ML-DSA signing cannot guarantee timely completion
even at the lowest security level.  Systems deploying ML-DSA on
\armmo{} should implement either:
(i)~pre-computation strategies that sign during idle periods, or
(ii)~timeout-and-retry mechanisms with appropriate fallback
behaviour.
Alternatively, hash-based signatures (SLH-DSA/FIPS~205)
offer deterministic signing times and merit investigation on this
platform.

\subsection{Device-Class Compatibility}
Based on our measurements, Table~\ref{tab:recommendation}
provides deployment recommendations for IETF RFC~7228
device classes.  As a reminder of the classification used
throughout this section: RFC~7228 Class~0 devices have
${\ll}\SI{10}{\kibi\byte}$ of RAM and ${\ll}\SI{100}{\kibi\byte}$
of code space, Class~1 approximately \SI{10}{\kibi\byte} and
\SI{100}{\kibi\byte}, and Class~2 approximately
\SI{50}{\kibi\byte} and \SI{250}{\kibi\byte}~\cite{rfc7228}.
The RP2040, with \SI{264}{\kibi\byte} of SRAM, exceeds all
three and is used here to measure what the algorithms cost, not
as a device under classification.

\begin{table}[t]
\centering
\caption{PQC deployment feasibility by RFC~7228 device
  class~\cite{rfc7228}, using the Total~RAM and code-size figures
  of Table~\ref{tab:memory} (PQClean reference~C).  Class
  boundaries are quoted verbatim from RFC~7228 Table~1
  (KiB $=$ 1024\,bytes).}
\label{tab:recommendation}
\footnotesize
\setlength{\tabcolsep}{3pt}
\begin{tabular}{>{\raggedright\arraybackslash}p{2.5cm}%
                >{\raggedright\arraybackslash}p{5.4cm}}
\toprule
\textbf{Class} & \textbf{Recommendation} \\
\midrule
Class~0 ($\ll$10\,KiB RAM,
  $\ll$100\,KiB Flash) & Not feasible for software PQC in the
  reference~C form measured here: ML-KEM-512 decapsulation alone
  needs 12.6\,KiB.  A hardware accelerator or a stack-reduced
  implementation is required. \\
Class~1 (${\sim}$10\,KiB RAM,
  ${\sim}$100\,KiB Flash) & Marginal.  ML-KEM-512 decapsulation
  (12.6\,KiB total RAM, 5.1\,KiB code) exceeds the nominal
  Class-1 data budget with reference~C; it becomes viable only
  with the stack-reduced variants reported for M4-class
  parts~\cite{kannwischer2019pqm4}.  No ML-DSA variant fits
  ($\geq$56.8\,KiB). \\
Class~2 (${\sim}$50\,KiB RAM,
  ${\sim}$250\,KiB Flash) & ML-KEM-512/768/1024 all fit
  (12.6--26.2\,KiB).  ML-DSA-44 (56.8\,KiB) exceeds the nominal
  budget; ML-DSA-65 and~-87 are out of reach. \\
Beyond Class~2 (e.g.\ RP2040,
  264\,KB SRAM) & All ML-KEM and ML-DSA variants fit.
  ML-DSA-44 is feasible with variance-aware scheduling; see
  Section~\ref{sec:results} for the latency distribution that
  such scheduling must accommodate. \\
\bottomrule
\multicolumn{2}{p{7.9cm}}{\footnotesize\vspace{2pt}
  \textbf{Scope of these recommendations.}  The memory rows
  transfer across \armmo{} implementations: stack and code size
  are properties of the compiled implementation, not of the
  vendor's multiplier choice.  The \emph{latency} judgements do
  not.  The RP2040 integrates the single-cycle multiplier, which
  ARMv6-M leaves optional; parts built with the 32-cycle
  iterative multiplier will be materially slower on the
  multiply-intensive NTT inner loops, and no timing conclusion
  here should be carried to such a part without re-measurement.
  Section~\ref{sec:multiplier} estimates that penalty at
  ${\approx}1.5\times$ from instruction counts; the memory rows
  are unaffected.}
\end{tabular}
\end{table}

\textbf{How far the map would move for optimised code.}
Table~\ref{tab:recommendation} is a reference~C map, and a reader
designing a real deployment will want to know whether an optimised
implementation changes the class assignment rather than merely the
margin.  We deliberately do \emph{not} answer this with a numerical
re-scaling of Table~\ref{tab:recommendation}.  An earlier version of
this analysis scaled the Total~RAM column by a single factor derived
from a vendor product announcement; we have removed that
extrapolation, because a figure quoted for one algorithm under
unstated measurement conditions is not a sound quantitative input to
a feasibility table covering six variants.  We cannot substitute a
measurement of our own: the stack figures in
Table~\ref{tab:memory} are PQClean reference~C, which is precisely
the baseline the extrapolation sought to improve on, and we did not
port a stack-reduced implementation to the RP2040.

What can be said without such a factor is directional, and it is
enough for a deployment decision.  Stack-reduced and
assembly-optimised implementations are known to require
substantially less peak RAM than PQClean reference~C---commercial
implementations advertise ML-KEM on Cortex-M in single-digit
kilobytes~\cite{pqshield2026}, and the stack-reduced variants
reported for M4-class parts~\cite{kannwischer2019pqm4} are well
below the reference~C figures---so optimisation moves variants
only \emph{down} the class table, never up.  Every feasibility
verdict in Table~\ref{tab:recommendation} is therefore a
conservative bound that remains valid for optimised code, and the
specific prize on offer is ML-KEM at class~1: our reference~C
ML-KEM-512 total of \SI{12.6}{\kilo\byte} sits close enough to the
\SI{10}{\kibi\byte} class-1 ceiling that a stack-reduced
implementation is a plausible route across it.  Establishing
whether it actually crosses requires measuring such an
implementation on ARMv6-M, which we identify as future work rather
than estimate here.

The same restraint applies to latency.  pqm4's optimised assembly
is \numrange{2}{4} times faster than reference~C on M4, but that
speedup draws on DSP and \texttt{UMULL} instructions the M0+
lacks, so it cannot be transplanted to ARMv6-M; we make no
predicted optimised handshake time.

\subsection{M0+ vs M4: Architectural Implications}
The measured 2.01--2.04$\times$ slowdown is considerably lower
than the 3--7$\times$ range sometimes cited in literature.
This result must be contextualised:
our comparison uses PQClean reference~C on M0+ versus pqm4
reference~C on M4, both at \texttt{-O3}, so the ratio is not
confounded by the optimisation level.

The compiler \emph{version}, however, is not matched, and this
paper's own findings say that matters.  pqm4 reports that the
figures in \texttt{benchmarks.csv} were produced with
\texttt{arm-none-eabi-gcc} (Arm GNU Toolchain 11.3.Rel1)
\texttt{11.3.1 20220712}, against our \texttt{12.2.1} (campaign~A)
and \texttt{15.2.1} (campaign~C).  Section~\ref{sec:interunit}
measures a \SIrange{1.4}{2.5}{\percent} shift in mean latency from
a compiler-version change alone at fixed \texttt{-O3}; applying
that as a two-sided bound to the M4 side widens the slowdown
envelope from \numrange{2.01}{2.04} to roughly
\numrange{1.96}{2.09}.  This is an unquantified residual in the
sense that we have not rebuilt pqm4 under our own toolchain to
measure it directly, and the bound assumes the M4 responds to a
version bump as the M0+ does.  The headline conclusion---a
slowdown near 2$\times$, not the 3--7$\times$ often assumed---is
unaffected, but the second decimal place should not be read as
significant across toolchains.  The replication campaign
quantifies what that choice is worth on this platform: rebuilding
the identical source at \texttt{-Os} costs 4.3--15.8\% in
latency (Table~\ref{tab:replication}), so a size-optimised M0+
build measured against this same \texttt{-O3} M4 baseline would
show a \emph{larger} slowdown, nearer 2.1--2.4$\times$: the M0+
side loses time while the reference does not move.  We state that
comparison in one direction only, because the pqm4 figures we
compare against are published at \texttt{-O3} and we have not
rebuilt them at \texttt{-Os}; a deployment that size-optimised both
sides would give up latency on the M4 as well, and the ratio would
fall back toward the flag-matched 2.01--2.04$\times$.  The
2.1--2.4$\times$ estimate is thus the penalty for size-optimising
the M0+ alone against a fixed reference, not a prediction for a
uniformly \texttt{-Os} system.
The M4 advantage would increase significantly with
assembly-optimised NTT implementations that exploit
\texttt{UMULL}, barrel shifts, and SIMD halfword operations.
Nevertheless, for the common case where IoT firmware uses
vendor-provided C libraries without custom assembly, our results
indicate that the M0+-to-M4 penalty is modest for ML-KEM.

\subsection{What Would a 32-Cycle Multiplier Cost?}
\label{sec:multiplier}
ARMv6-M leaves the multiplier to the licensee: the RP2040
integrates the single-cycle $32\times32\rightarrow32$ unit, while
many \armmo{} parts use the 32-cycle iterative alternative, i.e.\
\num{31} extra cycles per \texttt{MULS}.  Those parts are exactly
the sub-\$5 endpoints this paper's motivation invokes, so the
caveat deserves a number rather than only an acknowledgement.  What
follows is arithmetic on instruction counts, not a measurement, and
we label it as such throughout.

\textbf{The penalty is bounded by multiply density, and that
density is low.}  The extra cost falls only on multiply
instructions, so the slowdown is
\begin{equation}
S \;=\; 1 + 31\,\frac{M}{C},
\label{eq:multiplier}
\end{equation}
for $M$ multiply instructions executed over $C$ total cycles.  Two
facts keep $M/C$ small in reference~C.  First, the dominant cost on
this core is Keccak: \texttt{SHAKE128}/\texttt{256} drives matrix
expansion, noise sampling and hashing, and in \mldsa{} also mask
expansion and challenge derivation, and it consists of rotations,
XORs and ANDs with no multiply at all.  Second, the multiplies in
the lattice arithmetic can be counted from the implementation's
structure.

\textbf{Counting them for ML-KEM.}  PQClean's \texttt{fqmul(a,b)}
is one $16\times16\rightarrow32$ \texttt{MULS} followed by
\texttt{montgomery\_reduce}, which contributes two more (the
$q^{-1}$ product and the $\times q$ product): three multiplies per
modular multiplication.  A 256-coefficient NTT performs
$7\times128=896$ butterflies with one \texttt{fqmul} each
(${\approx}2.7\times10^{3}$ multiplies); the inverse transform adds
a scaling pass over all \num{256} coefficients
(${\approx}3.5\times10^{3}$ in total); \texttt{basemul} over a
polynomial pair is \num{128} blocks of five \texttt{fqmul}
(${\approx}1.9\times10^{3}$); \texttt{poly\_reduce} and
\texttt{poly\_tomont} contribute one multiply per coefficient.
ML-KEM-512 key generation runs four forward NTTs, four base
multiplications and the associated reductions, giving
$M\approx2\times10^{4}$ against the measured
$C=1.243\times10^{6}$ cycles (Table~\ref{tab:mlkem}): a multiply
share of ${\approx}\SI{1.6}{\percent}$, and by
(\ref{eq:multiplier}) a slowdown of ${\approx}1.5\times$.
Encapsulation and decapsulation add a re-encryption and further
transforms, but also proportionally more hashing, and land in the
same range.

\textbf{ML-DSA arrives at the same place by a different route.}
Its coefficients are 32-bit and the reference Montgomery reduction
consumes a 64-bit intermediate, which ARMv6-M must synthesise from
three to four \texttt{MULS} plus adds
(Section~\ref{sec:background}), so the per-coefficient multiply
cost is roughly double ML-KEM's.  The Keccak share rises in step,
however---every rejection-sampling iteration re-expands a mask and
re-derives a challenge---so the density stays in the low
single-digit percent and the estimate remains
${\approx}\numrange{1.5}{1.6}\times$.

\textbf{Sensitivity.}  Because (\ref{eq:multiplier}) is linear in
the multiply share, the estimate degrades gracefully rather than
collapsing: Table~\ref{tab:multiplier} evaluates $S$ over a range
that brackets any plausible counting error.  A threefold undercount
on our part still leaves the penalty below $2.5\times$, and the
\numrange{3}{7}$\times$ M0+/M4 figure sometimes quoted would
require multiplies to occupy \SIrange{6}{20}{\percent} of all
cycles---four to twelve times our estimate---which is
irreconcilable with a Keccak-dominated reference~C profile.

\textbf{What this changes.}  At $1.5\times$, an ML-KEM-512
handshake would take ${\approx}\SI{54}{\milli\second}$ on a
32-cycle-multiplier part: still comfortably feasible, and still
${\approx}11\times$ faster than our ECDH\,P-256 baseline even on
the conservative assumption that the classical baseline pays no
multiplier penalty at all (mbedTLS big-number arithmetic is
multiply-dense, so the true ratio would widen).  The \mldsa{}
verdict moves the other way: an \mldsa{}-44 $p_{99}$ of
\SI{584}{\milli\second} becomes ${\approx}\SI{0.9}{\second}$, so
the conclusion above---that \mldsa{} signing cannot meet a
\SI{100}{\milli\second} deadline on this class---holds with more
room to spare.  Memory figures are unaffected, being properties of
the compiled implementation rather than of the multiplier.  We
report all of this as an estimate to be tested and identify
replication on a 32-cycle-multiplier ARMv6-M part as future work;
it is the single measurement that would most extend the practical
reach of these results.

\begin{table}[t]
\centering
\caption{Estimated slowdown $S=1+31(M/C)$ on an \armmo{} part with
  the 32-cycle iterative multiplier, as a function of the share of
  cycles spent on multiply instructions.  Our instruction-count
  estimate for reference~C ML-KEM-512 is \SI{1.6}{\percent}
  (bold); the remaining columns bracket counting error.  This is
  arithmetic on (\ref{eq:multiplier}), not a measurement.}
\label{tab:multiplier}
\footnotesize
\setlength{\tabcolsep}{4pt}
\begin{tabular}{lrrrrr}
\toprule
\textbf{Multiply share (\%)} & 1 & \textbf{1.6} & 3 & 5 & 10 \\
\midrule
\textbf{Est.\ slowdown} & 1.31$\times$ & \textbf{1.50$\times$} &
  1.93$\times$ & 2.55$\times$ & 4.10$\times$ \\
\bottomrule
\end{tabular}
\end{table}

\subsection{RP2040-Specific Optimisation Opportunities}
Several RP2040-specific features could further reduce PQC
latency:
(i)~the second core could execute cryptographic operations
while the primary core handles I/O;
(ii)~the hardware interpolator co-processors could accelerate
address-generation patterns in the NTT;
(iii)~placing hot NTT loops in SRAM (via the
\texttt{\_\_not\_in\_flash\_func} attribute) could eliminate
QSPI flash access latency---the RP2040's XIP cache (16\,KB,
2-way set-associative) was active during all measurements, but
NTT working sets may exceed this cache, causing flash wait states.
These optimisations are left as future work.

\subsection{Communication Overhead Considerations}
For bandwidth-constrained IoT links, PQC object sizes matter as
much as computation time.  ML-KEM-512 requires transmitting an
800-byte public key and a 768-byte ciphertext (1,568\,bytes total
for a key exchange), compared to 65+65\,bytes for ECDH\,P-256.
This fits within a single CoAP/DTLS datagram
(${\sim}$1,280\,bytes MTU) with fragmentation, but exceeds typical
LoRa (222\,bytes) and BLE (244\,bytes) maximum payloads, requiring
application-layer fragmentation or multiple radio transmissions.
ML-DSA-44 signatures (2,420\,bytes) and public keys (1,312\,bytes)
are similarly challenging for constrained links.
Hybrid classical+PQ modes (e.g., ECDH+ML-KEM as recommended by
CNSA~2.0) would roughly double both the computation
(${\sim}$\SI{653}{\milli\second}) and bandwidth requirements,
though the PQ component dominates neither.

\subsection{Limitations}
\label{sec:limitations}
This study has several limitations that should be
acknowledged:
\begin{itemize}
  \item \textbf{Two units, one procurement batch}: campaigns~A
    and~B ran on a single RP2040; campaign~C
    (Section~\ref{sec:interunit}) adds a second unit and finds all
    \num{23} adequately powered operations equivalent within
    $\pm1\%$, with post-quantum operations agreeing to
    \SI{0.075}{\percent}.  This bounds inter-unit variation under
    realistic procurement, but both units were bought together and
    are likely to share a wafer lot, so it is not a
    silicon-variation study across process corners.  Temperature
    and supply-voltage dependence were not varied; all
    measurements were taken at laboratory ambient on USB power.
  \item \textbf{Extreme tail}: the \SI{10000}{}-signature
    replication campaign (Section~\ref{sec:replication}) resolves
    the distribution to the p99.9 quantile, but no finite sample
    bounds the worst case admitted by the standard.  Deployments
    with hard real-time deadlines require the
    worst-case-execution-time methodology of
    Riou et al.~\cite{pqshield2026pitfalls} rather than the
    percentile statistics reported here.  We report percentiles
    and the full raw distribution precisely so that the
    distinction is visible.
  \item \textbf{Hedged signing only}: all \mldsa{}
    measurements use the hedged variant with a 128-byte message
    over the Pure interface.  The deterministic variant and
    other message lengths were not measured; both shift absolute
    signing times, though neither changes the rejection
    structure.
  \item \textbf{Reference C only}: no ARMv6-M assembly
    optimisations were applied.  Optimised implementations
    (e.g., the multi-moduli NTT approach of
    Li et al.~\cite{li2023indocrypt}) would substantially
    reduce cycle counts.
  \item \textbf{Single-cycle multiplier}: every timing figure comes
    from a part carrying the optional single-cycle
    $32\times32\rightarrow32$ multiplier.
    Section~\ref{sec:multiplier} estimates the penalty on a
    32-cycle-multiplier ARMv6-M part at ${\approx}1.5\times$ from
    instruction counts, but we measured no such part; that estimate
    is analytic and untested.  Memory figures are unaffected.
  \item \textbf{No side-channel \emph{evaluation}, but one
    timing channel demonstrated}: we performed no power or
    electromagnetic analysis and no leakage assessment.  We
    note, however, that the iteration-recovery method of
    Section~\ref{sec:latticefit} is itself a timing side
    channel: it recovers a secret-dependent iteration count from
    an uninstrumented clock.  Section~\ref{sec:sidechannel}
    discusses what that quantity does and does not reveal and
    how it relates to published attacks on rejection sampling;
    we claim no key recovery and did not attempt one.
    Bos et al.~\cite{bos2021masking} have shown
    that masking Kyber on M0+ adds approximately $10\times$
    overhead.  Moreover, Wang et al.~\cite{wang2024keccak}
    demonstrated EM fault injection against ML-KEM and ML-DSA's
    Keccak permutation on \armmo{} with 89.5\% success rate,
    highlighting that unprotected implementations on this platform
    are vulnerable to physical attacks.
  \item \textbf{ROSC randomness}: the RP2040's ring oscillator
    is not cryptographically certified.  Production deployments
    would require a DRBG seeded from a certified source.
  \item \textbf{No protocol overhead}: network protocol costs
    (TLS, DTLS, CoAP) are not included; our measurements reflect
    isolated algorithm execution only.
  \item \textbf{Energy estimation}: energy values are derived
    from a constant-power datasheet model, not direct current
    measurement.  Because the model assumes constant power,
    every energy figure in this paper is the corresponding
    latency multiplied by a single constant
    (\SI{79.2}{\micro\joule\per\milli\second}) and therefore
    carries no information beyond Tables~\ref{tab:mlkem}
    and~\ref{tab:mldsa}.  We give the constant and the derived
    figures in Appendix~\ref{sec:appendix-energy} only as a unit
    conversion for battery-budget calculations and label them
    \emph{estimated} throughout; it should not be read as a measurement, and the
    relative energy claims it supports are exactly the relative
    latency claims.  Saarinen's \emph{pqps} measurements~\cite{saarinen2019pqps} on
    Cortex-M4 demonstrated over 50\% variation in average power
    draw across different cryptographic primitives, driven by
    differing instruction mixes.  Consequently, intra-family
    comparisons (e.g., ML-KEM-512 vs 768 vs 1024) are reliable
    since they share the same code structure, but cross-family
    comparisons (ML-KEM vs ECDH vs RSA) may carry larger error
    than the 10--20\% suggested by the datasheet tolerance alone.
\end{itemize}

\section{Conclusion}
\label{sec:conclusion}

This paper presented the first systematic benchmarks of the
NIST-standardised ML-KEM (FIPS~203) and ML-DSA (FIPS~204) on
ARM~\armmo{}, the most constrained 32-bit processor class
widely deployed in IoT.
On the RP2040 at \SI{125}{\mega\hertz}, ML-KEM-512 completes a full key
exchange in \SI{35.7}{\milli\second} at an estimated
\SI{2.83}{\milli\joule} (datasheet power model)---%
$17\times$ faster than a complete but unauthenticated
ECDH\,P-256 key agreement.
On \armmo{} silicon carrying the 32-cycle iterative multiplier
instead of the RP2040's single-cycle unit, an instruction-count
estimate puts that handshake at ${\approx}\SI{54}{\milli\second}$
and still ${\approx}11\times$ ahead of the same baseline
(Section~\ref{sec:multiplier}); that is the figure a reader
deploying on non-RP2040 ARMv6-M parts should carry away.
ML-DSA signing is feasible but exhibits 66.0--73.5\% coefficient of
variation due to rejection sampling: at ML-DSA-44, resolved over
\num{10000} signatures, $p_{99}$ is \SI{584}{\milli\second} and the
observed maximum \SI{1157}{\milli\second}, while the ML-DSA-65
and~-87 quantiles rest on $n{=}100$ and are lower bounds
($p_{99}\ge\SI{1125}{\milli\second}$ at ML-DSA-87).
The \armmo{} incurs only a ${\sim}2.0\times$ slowdown
(\numrange{2.01}{2.04}$\times$ cycle-matched;
\numrange{1.96}{2.09}$\times$ allowing for compiler-version
drift) relative to published \armfour{} results when both run
reference~C code.
Measuring a second physical device establishes that these figures
are properties of the part rather than of one die: all \num{23}
adequately powered operations are equivalent within a pre-declared
$\pm1\%$ margin, and the post-quantum operations agree to
\SI{0.075}{\percent}.  The dominant source of variation in
absolute latency is not the silicon but the build: a
compiler-version change at fixed \texttt{-O3} moves the same
operations by \SIrange{1.4}{2.5}{\percent}, and a change of
optimisation level by \SIrange{4.3}{15.8}{\percent}.  That is why
every absolute figure in this paper is reported with the build
that produced it.

These results demonstrate that lattice-based PQC is practical on
sub-\$5 IoT processors today, making the ``harvest now, decrypt
later'' threat addressable even for the most constrained device
class.  We release all code, data, and analysis scripts as an
open-source benchmark suite to enable reproducibility and
community extension.

Future work includes:
(i)~direct energy measurement with a shunt-resistor or INA219
current trace, replacing the datasheet unit conversion of
Appendix~\ref{sec:appendix-energy} with measured charge per
operation---the most repeated caveat in this paper, and the one
most likely to change a deployment decision, since duty-cycled
radios and sleep-mode residency make real consumption depart from a
constant-power model;
(ii)~ARMv6-M assembly optimisations for the NTT kernel;
(iii)~benchmarking SLH-DSA (FIPS~205) and FN-DSA (FIPS~206) on M0+;
(iv)~comparative evaluation on the RP2350 (Cortex-M33);
(v)~integration with DTLS\,1.3 and CoAP for end-to-end protocol
measurements;
(vi)~side-channel resistance evaluation; and
(vii)~replication on an \armmo{} part with the 32-cycle iterative
multiplier, testing the instruction-count estimate of
Section~\ref{sec:multiplier} against measurement.

\bibliographystyle{IEEEtran}


\appendices

\section{Estimated Energy per Operation}
\label{sec:appendix-energy}

This appendix previously carried a table of estimated energy per
operation.  Because that table was a re-scaling of the timing
results by a single constant and contained no measurement
independent of Tables~\ref{tab:mlkem} and~\ref{tab:mldsa}, we give
the constant and the derived figures in text instead.

The datasheet power model of Section~\ref{sec:methodology}
(\SI{3.3}{\volt}, \SI{24}{\milli\ampere},
\SI{79.2}{\milli\watt} constant active power) implies
\SI{79.2}{\micro\joule\per\milli\second}.  Multiplying any latency
in this paper by that constant gives the corresponding energy
estimate.  For the headline operations this yields, in mJ per
operation: \mlkem{} handshakes 2.83 (512), 4.48 (768) and 6.79
(1024); \mldsa{} sign cycles 19.22 (44), 31.46 (65) and 46.71
(87); and, for the classical baselines, 48.87 (ECDH\,P-256 key
agreement), 39.41 (ECDSA\,P-256 sign cycle) and 279.8 (RSA-2048
encrypt plus decrypt).

Two properties of these numbers should be kept apart.  Ratios
between them are robust, because the constant cancels: the
\SI{17}{\times} advantage of an \mlkem{}-512 handshake over the
ECDH\,P-256 baseline holds under any constant-power model.  The
absolute values are not robust---they inherit every inaccuracy of
the assumption that active power is constant and equal to the
datasheet figure---and no conclusion in this paper rests on them.
Direct current measurement is future work
(Section~\ref{sec:conclusion}).

\section{Goodness-of-Fit Checks for the Recovered Iteration
  Counts}
\label{sec:appendix-latticestats}

Section~\ref{sec:latticefit} reports a $\chi^2$ non-rejection of
the geometric model for the recovered iteration counts
($\chi^2=22.64$, 14~d.o.f., $p=0.066$, with $\hat{p}$ estimated
from the sample).  This appendix gives the three supporting
checks.

\textbf{The asymptotic approximation is not doing suspect work.}
No bin has an expected count below 59; the usual concern is
expected counts under~5.  A parametric bootstrap resampling
\num{20000} geometric datasets at $\hat{p}$ and recomputing the
statistic---estimating $\hat{p}$ afresh each time, so no
asymptotic assumption is used---gives $p=0.069$ against the
asymptotic \num{0.066}.

\textbf{A distribution-free check agrees.}  A
Kolmogorov--Smirnov test calibrated by the same bootstrap gives
$p=0.50$.  Its \emph{asymptotic} $p$-value is meaningless on
discrete data with an estimated parameter, and we do not quote
it.

\textbf{What the test could have detected.}  We take as the
alternative a beta-geometric family in which the per-attempt
acceptance probability drifts by a factor $\beta$ with each
successive attempt, so $\beta=1$ is the geometric null and
$\beta<1$ makes late attempts progressively less likely to
succeed.  At $n=\num{10000}$ and $\alpha=0.05$ the test reaches
\SI{80}{\percent} power at $\beta=0.988$, corresponding to
inflating $\mathbb{E}[k]$ by \SI{4.4}{\percent} (from \num{4.34}
to \num{4.53}); by $\beta=0.985$ (a \SI{5.8}{\percent} inflation)
power is \SI{96}{\percent}.  This is the basis for the bound
stated in the main text.

\textbf{Quantisation is not the explanation.}  A competing
account of the original rejection is that quantisation of the
early-abort step misassigns signatures between adjacent $k$
bins.  A synthetic control---\num{10000} signatures from an
exactly geometric process with cost and noise matched to the
campaign, run through the identical recovery procedure---recovers
the true $k$ for every signature at the observed noise level
(misassignment \SI{0}{\percent} against a residual standard
deviation of \SI{0.27}{\milli\second}), staying below
\SI{0.1}{\percent} at three times that noise and reaching
\SI{4.9}{\percent} only at ten times it.  It does not reproduce
the rejection, so we do not advance the explanation.

\textbf{The $t_{\min}$ anchor does not drive the result.}  The
affine model anchors $t_{\min}$ at the single fastest observed
signature, an extreme order statistic whose noise would otherwise
propagate into every recovered count.  Three checks on the
\num{10000}-signature campaign bound its influence.  Re-anchoring
on the 2nd through 100th-fastest signature changes $\mathbb{E}[k]$
by under \SI{0.1}{\percent} throughout; refitting with $t_{\min}$
free jointly with $(c,f)$ moves the anchor by
\SI{+0.010}{\milli\second} and $\mathbb{E}[k]$ by
\SI{0.07}{\percent}; and bootstrapping the minimum over
\num{2000} resamples gives a 95\% interval only
\SI{0.005}{\milli\second} wide, bounding the anchor's contribution
to the \SI{51.21}{\milli\second} setup estimate at
\SI{0.01}{\percent}.

The anchor is stable because it is not a lone draw from a diffuse
tail: \num{2346} of the \num{10000} signatures are
single-iteration and span only \SI{0.167}{\milli\second}, so the
sample minimum is the lower edge of a dense mode rather than an
outlier.  That is a property of this campaign and should not be
assumed at $n{=}100$---a further reason the recovery is claimed
only at \mldsa{}-44.  The script \texttt{tmin\_sensitivity.py}
reproduces all three checks.

\section{Multiplicity Control for the Inter-Unit Equivalence
  Tests}
\label{sec:appendix-multiplicity}

Section~\ref{sec:interunit} reports \num{23} equivalence verdicts.
Twenty-one carry a defined TOST $p$-value; the other two
(\texttt{randombytes} at \num{32} and \num{64} bytes) returned an
identical mean on both units with zero dispersion, so no test
statistic exists for them.  We exclude those two from the family
rather than count them as passing, while dividing $\alpha$ by the
full \num{23} claims rather than by the \num{21} tests---the
stricter of the two readings.

The largest TOST $p$-value among the \num{21} is \num{1.3e-7}
(RSA-2048 decryption), against a Bonferroni threshold of
$\alpha/23=\num{2.2e-3}$; the step-down Holm procedure rejects
every null as well.  The margin comes from the effect sizes rather
than from the sample sizes: the observed differences sit one to two
orders of magnitude inside an equivalence margin that was itself
pre-declared well below the toolchain drift the same hardware
exhibits (Section~\ref{sec:replication}).  Per-operation verdicts,
$p$-values and intervals are in
\texttt{results/campaign\_c/inter\_unit\_comparison.csv}.

\section{Benchmark Source Code}
\label{sec:appendix-code}

The complete benchmark source code, raw data, analysis scripts,
and reproduction instructions are available in the companion
open-source repository:

\begin{center}
\url{https://github.com/rojinc/pqc-cortex-m0-benchmark}
\end{center}

\noindent The repository includes:
\begin{itemize}
  \item \texttt{src/bench\_harness.h} --- timing and stack
    measurement framework using the Pico~SDK's
    \texttt{time\_us\_32()} function;
  \item \texttt{src/bench\_mlkem.c} --- ML-KEM benchmarks for
    all three security levels (30 runs each);
  \item \texttt{src/bench\_mldsa.c} --- ML-DSA benchmarks with
    100-run signing for variance analysis;
  \item \texttt{src/bench\_classical.c} --- RSA-2048, ECDSA\,P-256,
    and ECDH\,P-256 baselines using mbedTLS~3.6.0;
  \item \texttt{scripts/add\_energy.py} --- datasheet-based energy
    estimation post-processor;
  \item \texttt{scripts/reproduce.py} --- environment fingerprinting
    and cross-run statistical comparison;
  \item \texttt{results/raw/} --- raw CSV benchmark data for
    campaign~A;
  \item \texttt{results/campaign\_b/} --- the independent
    replication campaign of Section~\ref{sec:replication}
    ($n{=}1000$ deterministic, $n{=}\num{10000}$ signing);
  \item \texttt{results/campaign\_c/} --- the inter-unit campaign
    of Section~\ref{sec:interunit}: per-device raw CSVs for both
    RP2040 units with a metadata sidecar per capture recording
    the hardware unique ID, \texttt{clk\_sys}, firmware build
    stamp, and row reconciliation, plus the per-operation
    verdict table \texttt{inter\_unit\_comparison.csv};
  \item \texttt{scripts/capture\_campaign\_c.py} --- serial
    capture with row-count reconciliation, which exits non-zero
    on a dropped row or a truncated capture rather than writing a
    short file;
  \item \texttt{scripts/compare\_boards.py} --- the TOST
    equivalence analysis of Section~\ref{sec:interunit},
    including the minimum-$n$ power requirement and the
    distribution-free corroboration test that a difference
    verdict must clear;
  \item \texttt{scripts/verify\_campaign\_b.py} --- recomputes
    every replication, tail, and lattice-fit value reported here
    directly from the raw data, exiting non-zero on any mismatch;
  \item \texttt{scripts/verify\_revision.py} --- recomputes every analysis
        added during peer review: the identifiability result of
        Section~\ref{sec:latticegen}, the corrected $\chi^2$ test of
        Section~\ref{sec:latticefit}, the bootstrap, power analysis
        and synthetic control of
        Appendix~\ref{sec:appendix-latticestats}, the
        quantisation bound on the reported intervals, the
        confidence intervals of Table~\ref{tab:replication}, the
        multiply-count estimate of Section~\ref{sec:multiplier},
        and the \armfour{} comparison of Table~\ref{tab:m0m4}.  A
        \texttt{-{}-fast} flag skips the bootstrap-heavy first section;
        the script exits non-zero on any mismatch;
  \item \texttt{REPRODUCE.md} --- step-by-step reproduction guide.
\end{itemize}

\end{document}